\documentclass[a4paper,fleqn]{cas-sc}

\usepackage[authoryear,longnamesfirst]{natbib}
\usepackage{amsmath,amssymb,amsfonts}
\usepackage{graphicx}      
\usepackage{textcomp}
\usepackage{xcolor}
\usepackage{stfloats} 
\usepackage{float}
\usepackage{accents}
\usepackage{tikz}
\usetikzlibrary{calc}
\usetikzlibrary{shapes, arrows,automata}
\usetikzlibrary{shapes.multipart}
\usepackage[latin1]{inputenc}   
\usepackage{dsfont}

\usepackage{enumitem}
\usepackage{subfigure}
\usepackage{caption}
\usepackage{algorithm}
\usepackage{listofitems}
\usepackage{pgfplots}
\usepackage{multirow}
\usepackage{url}

\pgfplotsset{compat=1.18}
\usepackage{array}
\newcolumntype{P}[1]{>{\centering\arraybackslash}p{#1}}

\newtheorem{theorem}{Theorem}
\newtheorem{definition}{Definition}
\newtheorem{assumption}{Assumption}

\newtheorem{remark}{Remark}

\usepackage{xcolor}

\tikzstyle{sum} = [draw, fill=blue!20, circle, node distance=1cm]
\tikzstyle{input} = [coordinate]
\tikzstyle{pinstyle} = [pin edge={to-,thin,black}]

\usetikzlibrary{tikzmark, positioning, fit, shapes.misc}
\usetikzlibrary{decorations.pathreplacing, calc}
\tikzset{brace/.style={decorate, decoration={brace}},
    brace mirrored/.style={decorate,decoration={brace,mirror}},
}

\newsavebox{\DNN}
\usepackage{listofitems} 
\tikzstyle{mynode}=[thick,draw=blue,fill=blue!30,circle,minimum size=25]

\definecolor{GG}{rgb}{0,0,1}

\usepackage{tabularx}
\usepackage{longtable} 
\usepackage{colortbl}
\usepackage[thinlines]{easytable}

\newcommand{\smallmat}[1]{\left[ \begin{smallmatrix}#1 \end{smallmatrix} \right]}

\DeclareMathAlphabet\mathbfcal{OMS}{cmsy}{b}{n}

\usepackage{transparent}

\usepackage{adjustbox} 

\usepackage{pdfpages} 
\usepackage{afterpage}
\usepackage{lipsum} 
\usepackage{fancyhdr} 
\usepackage{color} 
\usepackage{soul} 
\usepackage{eurosym}  
\usepackage{algpseudocode}

\def\tsc#1{\csdef{#1}{\textsc{\lowercase{#1}}\xspace}}
\tsc{WGM}
\tsc{QE}
\tsc{EP}
\tsc{PMS}
\tsc{BEC}
\tsc{DE}

\begin{document}
\let\WriteBookmarks\relax
\def\floatpagepagefraction{1}
\def\textpagefraction{.001}
\shorttitle{Constrained Performance Boosting Control for Nonlinear Systems}
\shortauthors{G. Giacomelli et~al.}

\title [mode = title]{Constrained Performance Boosting Control for Nonlinear Systems}                      
\tnotemark[1]

\tnotetext[1]{This work was supported by NCCR Automation, grant agreement 51NF40\textunderscore225155 from the Swiss National Science Foundation.}


\author[1]{Gianluca Giacomelli}[orcid=0009-0003-9008-493X]

\ead{g.giacomelli@tue.nl}

\author[2]{Danilo Saccani}[orcid=0000-0002-3059-9160]
\ead{danilo.saccani@epfl.ch}

\author[1]{Siep Weiland}[orcid=0000-0002-6131-4658]
\ead{s.weiland@tue.nl}

\author[2]{Giancarlo Ferrari-Trecate}[orcid=0000-0002-9492-9624]
\ead{giancarlo.ferraritrecate@epfl.ch}

\author[1]{Valentina Breschi}[orcid=0000-0002-1533-7349]
\ead{v.breschi@tue.nl}


\affiliation[1]{organization={Control Systems Group, Eindhoven University of Technology},
                city={Eindhoven},
                postcode={5612AZ}, 
                country={The Netherlands}}

\affiliation[2]{organization={Institute of Mechanical Engineering, \'{E}cole Polytechnique F\'{e}d\'{e}rale de Lausanne (EPFL)},
                postcode={CH-1015}, 
                city={Lausanne},
                country={Switzerland}}



\begin{abstract}
We present the Alternating Direction Method of Multipliers (ADMM) for Performance Boosting (PB), an approach for designing neural controllers for stable nonlinear systems subject to state and input constraints. The method builds on an internal model control formulation of PB. In this setting, the controller is parametrized as a stable neural operator, so closed-loop stability is guaranteed by construction, and its weights are trained offline to improve performance. 
To provide a systematic procedure for promoting constraint satisfaction during training, we reformulate the finite-horizon problem of the PB formulation by introducing auxiliary state and input trajectories. This augmentation allows us to cast an ADMM-based algorithm that alternates between two steps: a gradient-descent-based update of the controller parameters, having the same structure as the PB training problem without explicit constraints, and a projection step that promotes the trajectory feasibility. As a result, this procedure handles constraints during training without altering the controller architecture or compromising its stability-by-design guarantees. Indeed, the stability guarantee follows from the chosen stable controller parametrization, which is not changed in our framework with respect to the foundational PB formulation, and is independent of ADMM convergence. At the same time, this closed-loop stability guarantee does not imply performance optimality or closed-loop constraint satisfaction, which depend on the convergence of ADMM-PB, which is not yet guaranteed in this work.
Our numerical results show that, compared with a baseline based on barrier-inspired soft penalties in the loss, ADMM-PB achieves lower constraint violations, at the price of more conservative closed-loop behavior.
\end{abstract}




\maketitle

\section{Introduction}\label{sec:Intro}
The increasing demand for high-performing controllers calls for the design of control architectures that improve system performance while still ensuring closed-loop stability~\citep{Wang2023}. 
At the same time, operational limits and safety-critical applications require these architectures to prioritize constraint satisfaction~\citep{Xiao2024}. These challenges are further heightened by the growing complexity of modern dynamical systems, which calls for increasingly flexible control architectures. In this context, Neural-Network (NN) controllers have shown impressive empirical performance on complex control tasks, owing to their ability to approximate rich and highly nonlinear feedback policies~\citep{kaufmann2023champion,lee2020learning}. Yet, such flexibility typically comes at the cost of limited formal guarantees on the resulting closed-loop behavior. 
Within this challenging landscape, recent works have proposed neural parameterizations that characterize all and only the stability-preserving controllers for stable or pre-stabilized nonlinear systems~\citep{furieri2024learning,galimberti2025parametrizations} based on Internal Model Control (IMC)~\citep{garcia1982internal,economou1986internal} and nonlinear Youla parameterizations~\citep{fujimoto2000characterization,paice2002class}.
Such parameterizations have been used for Performance Boosting (PB) (see~\cite{furieri2024learning,galimberti2025parametrizations}), yielding strategies that enable a clean separation between stability, enforced by construction through the controller architecture, and performance, which is optimized over the trainable parameters. 
These 
approaches  
amount to solving \emph{offline} an unconstrained Nonlinear Optimal Control (NOC) problem over the controller parameters, with a loss that quantifies the desired boosting objective. However, although this formulation improves performance while preserving \emph{closed-loop stability by design}, it is not explicitly tailored to enforce state and input constraints. In practice, constraints can only be promoted through loss augmentations \cite[Section VI.B]{furieri2024learning}, e.g., via ad hoc barrier-inspired penalty terms, rather than handled through a systematic mechanism.

Different approaches are available to move beyond barrier-inspired penalties for constraint handling when controlling a nonlinear system, which can be distinguished into two macro-classes. 
A first class of approaches enforces them \emph{online}, by recomputing or correcting the control action at runtime, e.g., as done by reference governors~\citep{garone2017} and predictive safety filters~\citep{wabersich2021predictive}. However, these approaches introduce an additional layer in the control architecture and, in many cases, require an online optimization or feasibility computation, thereby increasing implementation complexity. Instead, a second class seeks to account for constraints directly during controller synthesis or training, e.g., through Lyapunov- or barrier-based design~\citep{lin1991universal, ames2019control}, eventually by incorporating such certificates or penalties in the training objective and then verifying the resulting properties a posteriori~\citep{dawson2023safe}. While this class of constraint-handling strategies aligns with the offline nature of the approach presented in~\cite{furieri2024learning}, these synthesis-oriented methods have two main drawbacks. On the one hand, they often rely on the availability of suitable Lyapunov or barrier certificates, whose design for general nonlinear systems can be highly nontrivial. Meanwhile, when simplified or conservative certificates are used, synthesis-oriented approaches may induce overly conservative closed-loop behavior, hampering performance. Nonetheless, since~\cite{furieri2024learning} already formulates controller design as the \emph{offline} training of a neural \emph{stable} operator, this raises the question of whether constraint-handling strategies specifically developed for neural policies can be incorporated during training, while preserving the original stability-preserving architecture. 
A first option is post-hoc certification, for instance via mixed-integer verification methods such as~\cite{karg2019learning}. Although rigorous, these approaches are computationally demanding and often scale poorly, making them less appealing when constraints should be incorporated already at design time. A second option is to use certificate-based safe design methods, such as the one proposed by~\cite{berkenkamp2017safe}, which ensure that the closed-loop evolution remains within a Region of Attraction (RoA). This guarantee may imply constraint satisfaction when the admissible set lies within that region. Yet, in nonlinear settings, constructing suitable certificates and nonconservative RoA estimates is challenging, and conservative approximations may severely restrict the set of admissible policies. Finally, one may enforce constraints directly at the architecture level through projection layers~\citep{Morari,paulson2020approximate,grontas2025pinet}. While this approach is well-suited to handle input constraints, extending it to state constraints requires accounting for the system dynamics. In a nonlinear setting, this might nonetheless lead to projections that are no longer convex nor easily computable.

With the aim of preserving the control architecture proposed in~\cite{furieri2024learning}, in this work, we extend its learning procedure to account for convex state and input constraints. More precisely, we seek to handle these constraints without modifying the underlying stable-by-design controller parametrization, the closed-loop rollout\footnote{A rollout is the procedure of updating the NN parameters through gradient descent after substituting the input-output data generated by the closed-loop dynamics in the training loss.} used in training, or the separation between stability and performance characterizing~\cite{furieri2024learning}. To this end, we introduce ADMM-PB, a learning procedure based on the Alternating Direction Method of Multipliers (ADMM)~\citep{boyd2011distributed} to solve a constrained version of the NOC arising in PB. It is worth remarking that ADMM has already been explored for neural-network training with encouraging empirical results as an alternative to gradient-based methods~\citep{guan2021pdladmm,taylor2016training}. However, our goal in this work is not to modify the rollout-based structure of the original training step in~\cite{furieri2024learning}, but rather to use ADMM to account for state and input constraints during training. 
It is also worth noting that available convergence results for nonconvex settings typically rely on lifted network formulations, relaxations, or additional structural assumptions. These assumptions do not directly fit the nonlinear rollout coupling or the stable-by-design controller parameterization considered here (see, e.g., the assumptions in~\cite{wang2019admm}). 
Accelerated primal-dual methods could, in principle, also be considered. However, these approaches do not allow for the same clean separation between neural parameter optimization and trajectory projection onto the feasible set as ADMM. Moreover, they are primarily designed for convex formulations~\citep{zhao2024accelerated,zhao2023accelerated}, requiring tailored extensions to handle the non-convex nature of our performance-boosting problem.

Hence, while ADMM is not the only primal-dual method to achieve our goal, it is particularly convenient for the constrained PB formulation considered here. Indeed, after introducing auxiliary copies of the state and input trajectories, the training problem naturally decomposes into two blocks: a parameter update, which preserves the same rollout-based structure as the original unconstrained PB training step, and a feasibility step that reduces to Euclidean projections onto the admissible convex state and input sets. 
This splitting is particularly appealing because it decouples the different objectives we aim to achieve with the proposed procedure. The parameter update acts only on the weights of a controller that remains in the chosen stable-by-design parametrized class. Hence, the corresponding closed-loop stability guarantee is preserved at every ADMM-PB iteration and does not depend on the convergence of the ADMM routine. By contrast, constraint satisfaction is tied to the feasibility of the ADMM training problem. If the ADMM iterations converge and the primal residuals between rollout trajectories and auxiliary projected trajectories vanish, then the constraints are satisfied across the sampled training trajectories. If ADMM-PB is stopped early, stability guarantees are still retained, while feasibility and performance depend on the achieved residuals and on the quality of the trained solution. In this sense, ADMM thus provides a natural way to separate performance-oriented parameter training, projection-based constraint handling, and stability preservation.

\emph{Outline:} We first formally introduce the considered setting and the problem we aim to solve in Section~\ref{sec:background}. Then, the proposed solution (ADMM-PB) is outlined in Section~\ref{Sec:Approach}, while Section~\ref{sec:practical_aspects} provides some practical guidelines on the choices of a set of key hyperparameters of ADMM-PB. The effectiveness of the proposed approach is lastly analyzed in Section~\ref{sec:numerical_example} on a couple of benchmark examples. The paper ends with some conclusions and directions for future work. 

\emph{Notation:} We denote the set of natural numbers including zero and the set of real numbers as $\mathbb{N}_{0}$ and $\mathbb{R}$, respectively, and an index set as $[n_{el}]=\{1,\dots,n_{el}\}$ with $n_{el}\in\mathbb{N}$. Given a vector $b \in \mathbb{R}^{n_b}$, we denote its transpose as $b^{\top}$, and given two vectors $a \in \mathbb{R}^{n_a}$ and $b \in \mathbb{R}^{n_b}$, we denote with $\mathrm{col}(a,b)$ the column vector stacking them. For a given set $\mathcal{B} \subseteq \mathbb{R}^{n_b}$, the indicator function $\mathcal{I}_{\mathcal{B}}: \mathbb{R}^{n_b} \rightarrow \mathbb{R}$ is given by
\begin{equation*}\label{eq:indicator_function}
\mathcal{I}_{\mathcal{B}}(b)=\begin{cases}
    0, ~~~~~\mbox{ if } b \in \mathcal{B},\\
    +\infty, ~\mbox{ otherwise},
\end{cases}    
\end{equation*}
\textcolor{black}{while $\Pi_{\mathcal{B}}: \mathbb{R}^{n_b} \rightarrow \mathcal{B}$ denotes the projection operator, i.e.,}
\begin{equation*}\label{eq:projection}
    \textcolor{black}{\Pi_{\mathcal{B}}(b) = \underset{b^\pi\in\mathcal{B}}{\mathrm{argmin}} ~\|b^\pi- b\|_2^2.}    
\end{equation*}
Furthermore, we denote with $\mathcal{B}^{N}$ the cartesian product of the set $\mathcal{B}$ with itself $N$-times. Identity matrices and vectors of zeros are indicated as $I$ and $\boldsymbol{0}$, without specifying their dimensions. Given a signal $z_t \in \mathbb{R}^{n_z}$, with $t \in \mathbb{N}_0$, we define $z_{[t_1,t_2]}=\smallmat{
    z_{t_1}^{\top} & z_{t_1+1}^{\top} & \cdots & z_{t_2}^{\top}
}^{\top}$ for $t_1,t_2 \in \mathbb{N}_0$ such that $t_1<t_2$. Meanwhile, $\mathbf{z}=(z_{0},z_{1},\ldots) \in \ell^{n_z}$ denotes the sequence of values taken by $z_t$ for all $t\geq 0$. Furthermore, by introducing the $p$-norm of $\mathbf{z}$ as
\begin{equation*}  \|\mathbf{z}\|_{p}=\left(\sum_{t=0}^{+\infty}|z_t|^{p}\right)^{\!\!\frac{1}{p}} \text{for} \ p\in[1,\infty), \ \|\mathbf{z}\|_{\infty }= \sup_{t\geq0} |z_t|,
\end{equation*}
we say that $\mathbf{z} \in \ell_{p}^{n_z}\subset \ell^{n_z}$ when $\|\mathbf{z}\|_{p}<\infty$. According to this definition, we can formally introduce the definitions of $\ell_p$-stable operator with finite $\mathcal{L}_p$ gain as follows (see \cite{furieri2024learning}).
\begin{definition}\label{def:ellp_stable_operators} An operator $\mathbfcal{A}: \ell^{n_z} \mapsto \ell^{n_v}$ is said to be $\ell_p$-stable if it is causal and $\mathbfcal{A}(\mathbf{z}) \in \ell_p^{n_v}$ for all $\mathbf{z} \in \ell_{p}^{n_z}$; compactly we say $\mathbfcal{A}\in\mathcal{L}_p$. 
\end{definition}

\section{Setting and goals}\label{sec:background}
Consider a nonlinear, time-varying system whose dynamics are described by the difference equation
\begin{equation}\label{eq:state_space}
    x_{t}=f_t(x_{[0,t-1]},u_{[0,t-1]})+w_t,
\end{equation}
where $x_t \in \mathbb{R}^{n}$ and $u_t \in \mathbb{R}^{m}$ denote the system's state and the controlled input at time $t \in \mathbb{N}_0$, respectively, with $f_0(\cdot)=\boldsymbol{0}$. Meanwhile, $w_t \in \mathcal{W}_t \subseteq \mathbb{R}^{n}$ is a process noise realization with known distribution $\mathcal{D}_t$ (i.e., $w_t \sim \mathcal{D}_t$) and with $w_0=x_0$. Let us also consider the associated operator form, i.e.,
\begin{subequations}\label{eq:operator_form}
   \begin{equation} 
    \mathbf{x}=\mathbf{F}(\mathbf{x},\mathbf{u})+\mathbf{w},
\end{equation}
where $\mathbf{x}=(x_0,x_1,\ldots) \in \ell^{n}$, $\mathbf{u}=(u_0,u_1,\ldots) \in \ell^{m}$,  $\mathbf{w}=(x_0,w_1,\ldots)\in \ell^{n}$, and $\mathbf{F}:\ell^{n}\times \ell^{m} \rightarrow \ell^{n}$ is the strictly causal operator induced by the state dynamics, i.e., 
\begin{equation}
    \mathbf{F}(\mathbf{x},\mathbf{u})\!=\!(\boldsymbol{0},f_1(x_0,u_0),\ldots,f_{t}(x_{[0,t-1]},u_{[0,t-1]}),\ldots).
\end{equation} 
\end{subequations}
Since \eqref{eq:operator_form} produces a unique state sequence $\mathbf{x} \in \ell^{n}$ for a given $\mathbf{w} \in \ell^{n}$ and $\mathbf{u} \in \ell^{m}$, there exists a unique transition operator 
\begin{equation}\label{eq:transition_operator}
    \mathbfcal{F}: (\mathbf{u},\mathbf{w}) \mapsto \mathbf{x},
\end{equation}
characterizing the input-to-state behavior of the system, which we assume satisfies the following\footnote{Assumption~\ref{assumption:L_p_map} covers both open-loop $\ell_p$-stable and pre-stabilized plants.}(see Definition~\ref{def:ellp_stable_operators} in \cite{furieri2024learning}).
\begin{assumption}\label{assumption:L_p_map}
    The transition operator $\mathbfcal{F}$ in \eqref{eq:transition_operator} belongs to $\mathcal{L}_p$, i.e., $\mathbfcal{F} \in \mathcal{L}_p$.
\end{assumption}
Under Assumption~\ref{assumption:L_p_map}, we aim to design a nonlinear, state-feedback, time-varying control policy
\begin{equation}\label{eq:controller}
    \mathbf{u}=\mathbf{K}(\mathbf{x})=(K_0(x_0),K_1(x_{[0,1]}), \ldots, K_{t}(x_{[0,t]}),\ldots),
\end{equation}
with $\mathbf{K}: \ell^{n} \rightarrow \ell^{m}$ being the causal operator to be designed. 
By introducing the closed-loop operators $\boldsymbol{\Phi}^{\mathbf{x}}[\mathbf{F},\mathbf{K}]: \mathbf{w} \mapsto \mathbf{x}$ and $\boldsymbol{\Phi}^{\mathbf{u}}[\mathbf{F},\mathbf{K}]: \mathbf{w} \mapsto \mathbf{u}$ such that
\begin{equation}\label{eq:closed_loop_operators}
    \mathbf{x}=\boldsymbol{\Phi}^{\mathbf{x}}[\mathbf{F},\mathbf{K}]( \mathbf{w}),\qquad \mathbf{u}=\boldsymbol{\Phi}^{\mathbf{u}}[\mathbf{F},\mathbf{K}](\mathbf{w}),
\end{equation}
we formalize closed-loop $\ell_p$-stability as follows.
\begin{definition}[Closed-loop $\ell_p$-stability]\label{def:closed_loop_stab}
Given the system operator $\mathbf{F}$ in~\eqref{eq:operator_form}, a causal controller $\mathbf{K}: \ell^n \to \ell^m$ yields to a $\ell_p$-stable closed-loop system if the corresponding closed-loop operators in~\eqref{eq:closed_loop_operators} satisfy
\begin{equation}\label{eq}
\boldsymbol{\Phi}^{\mathbf{x}}[\mathbf{F},\mathbf{K}] \in \mathcal{L}_p,
\qquad
\boldsymbol{\Phi}^{\mathbf{u}}[\mathbf{F},\mathbf{K}] \in \mathcal{L}_p.
\end{equation}
\end{definition}
Our ideal objective is then to $(i)$ boost the performance of the considered system and $(ii)$ preserve $\ell_{p}$ stability, while $(iii)$ accounting for user-defined input and state constraints.

We can exploit the framework introduced in \cite[Problem 1]{furieri2024learning} to express this ideal constrained design objective as the following finite-horizon NOC problem:
\begin{subequations}\label{eq:NOC1}
    \begin{align}
        & \underset{\mathbf{K}(\cdot)}{\mathrm{minimize}}~~~\mathbb{E}_{w_{[0,T]}}[L(x_{[0,T]},u_{[0,T]})] \label{eq:NOC1_cost}\\
        &\qquad~~ \mbox{s.t. }~~x_{t}\!=\!f_{t}(x_{[0,t-1]},u_{[0,t-1]})\!+\!w_t,~\forall t \!\in\! [1,T],\\
        &\qquad\qquad~~~ w_0=x_0,\\
        &\qquad\qquad~~~ u_{t}=K_{t}(x_{[0,t]}), ~~~~\forall t \!\in\! [0,T], \label{eq:input_dynamics}\\
        &\qquad\qquad~~~ x_{t} \!\in\! \mathcal{X},~~u_{t} \!\in\! \mathcal{U}, ~~~~\forall t \!\in\! [0,T],\\
        &\qquad\qquad~~~ (\boldsymbol{\Phi}^{\mathbf{x}}[\mathbf{F},\mathbf{K}], \boldsymbol{\Phi}^{\mathbf{u}}[\mathbf{F},\mathbf{K}]) \in \mathcal{L}_{p}, \label{eq:NOC1_stability}
    \end{align}
\end{subequations}
where $L: (\mathbb{R}^{n} \times \mathbb{R}^{m})^{T+1} \rightarrow \mathbb{R}$ is the user-defined performance boosting loss, here assumed to be chosen to be piecewise differentiable and lower bounded, while $\mathcal{X}\subseteq \mathbb{R}^{n}$ and $\mathcal{U}\subseteq \mathbb{R}^{m}$ are prescribed \emph{convex} admissible sets for the closed-loop states and inputs, respectively. 
Using the IMC architecture (see \cite{garcia1982internal}) schematized in \figurename{~\ref{fig:Control_Scheme}}, the stability enforcing constraint in \eqref{eq:NOC1_stability} can be recast as a stability requirement on a learnable operator $\mathbfcal{M}: \mathbf{w} \mapsto \mathbf{u} $ by leveraging the following theorem.
\sbox{\DNN}{%
\scalebox{.15}{\begin{tikzpicture}[x=2.2cm,y=1.4cm]
  \readlist\Nnod{2,4,3,2} 
  \foreachitem \N \in \Nnod{ 
    \foreach \i [evaluate={\x=\Ncnt; \y=\N/2-\i+0.5; \prev=int(\Ncnt-1);}] in {1,...,\N}{ 
      \node[mynode] (N\Ncnt-\i) at (\x,\y) {};
      \ifnum\Ncnt>1 
        \foreach \j in {1,...,\Nnod[\prev]}{ 
          \draw[thick] (N\prev-\j) -- (N\Ncnt-\i); 
        }
      \fi 
    }
  }
\end{tikzpicture}}
}
\begin{figure}
\centering
\scalebox{1}{\begin{tikzpicture}
    \node[coordinate] (start) {};
    \node[draw,rectangle,below of=start,node distance=1cm,minimum height=1cm,minimum width=2cm,fill=blue!5!white] (Map) {\usebox{\DNN}};
    \node[draw,rectangle,right of=start,node distance=4cm,minimum height=1cm,minimum width=2cm, fill=blue!5!white] (sys) {$\mathbf{F}(\mathbf{x}, \mathbf{u})$};
    \node[draw,rectangle,left of=Map,node distance=4cm,minimum height=1cm,minimum width=2cm, fill=blue!5!white] (Model) {$\mathbf{F}(\mathbf{x}, \mathbf{u})$};
    \node[draw,circle,right of=sys,node distance=2cm,minimum height=.30cm] (Sum1){};
    \node[draw,circle,right of=Sum1,node distance=.4cm,inner sep=0cm, minimum size=1mm, fill=black] (Diram1){};
    \node[draw,circle,below of=Diram1,node distance=1cm,inner sep=0cm, minimum size=1mm, fill=black] (Diram2){};
    \node[draw,circle,right of=Model,node distance=2cm,minimum height=.30cm] (Sum2){};
    \node[draw=blue, dashed, rectangle,right of=Sum2,node distance=0cm,minimum height=3.5cm,minimum width=8cm] (Rec_C){};
    \node[above of=Rec_C,node distance=2cm, text=blue] (Text_C){Controller};

    \node[coordinate,left of=sys,node distance=1cm] (aid_1_sys) {};
    \node[coordinate,above of=aid_1_sys,node distance=.25cm] (Input_sys_1) {};
    \node[coordinate, below of=aid_1_sys,node distance=.25cm] (Input_sys_2) {};
    \node[coordinate,left of=Diram2,node distance=4cm] (aid_2_sys){};
    \node[coordinate,left of=Input_sys_2,node distance=.6cm] (aid_3_sys){};
    \node[draw,circle,left of=Input_sys_1,node distance=1.2cm,inner sep=0cm, minimum size=1mm, fill=black] (Diram0){};

    \node[coordinate,right of=Map,node distance=1.8cm] (aid_1_map) {};
    \node[coordinate,right of=Map,node distance=1cm] (aid_2_map) {};
    \node[coordinate,below of=aid_2_map,node distance=.6cm] (aid_3_map) {};
    \node[coordinate,left of=Map,node distance=1cm] (aid_4_map) {};
    \node[coordinate,below of=aid_4_map,node distance=.6cm] (aid_5_map) {};

    \node[coordinate,left of=Model,node distance=1cm] (aid_1_mod) {};
    \node[coordinate,above of=aid_1_mod,node distance=.25cm] (Input_mod_1) {};
    \node[coordinate, below of=aid_1_mod,node distance=.25cm] (Input_mod_2) {};
    \node[coordinate,left of=Input_mod_1,node distance=.6cm] (aid_2_mod){};
    \node[coordinate,left of=Input_mod_2,node distance=.6cm] (aid_3_mod){};
    \node[coordinate,above of=aid_2_mod,node distance=1cm] (aid_4_mod){};
    \node[coordinate,below of=aid_3_mod,node distance=1cm] (aid_5_mod){};
    \node[draw,circle,right of=aid_5_mod,node distance=3.6cm,inner sep=0cm, minimum size=1mm, fill=black] (Diram3){};
    \node[coordinate,right of=Diram3,node distance=8.4cm] (aid_6_mod){};
    
    \node[coordinate,above of=Sum1,node distance=.8cm] (Noise_start) {};
    \node[coordinate,right of=Sum1,node distance=1cm] (State_end) {};

    \draw[->]       (sys)   -- node[xshift=.15cm ,yshift=-.25cm]{+}(Sum1);
    \draw[->]       (Noise_start)   -- node[xshift=-.25cm ,yshift=.25cm]{$\mathbf{w}$} node[xshift=.25cm ,yshift=-.2cm]{+}(Sum1);
    \draw[->]       (Sum1)   -- node[xshift=.5cm ,yshift=.25cm]{$\mathbf{x}$}(State_end);
    \draw[-]       (Diram1)   -- (Diram2);
    \draw[-]       (Diram2)   -- (aid_2_sys);
    \draw[-]       (aid_2_sys)   -- (aid_3_sys);
    \draw[->]       (aid_3_sys)   -- (Input_sys_2);
    \draw[->]       (Diram0)   -- node[yshift=.25cm]{$\mathbf{u}$}(Input_sys_1);
    \draw[-]        (Map) -- (aid_1_map);
    \draw[-]        (aid_1_map) -- (Diram0);
    
    \draw[->]       (aid_2_mod)   -- (Input_mod_1);
    \draw[->]       (aid_3_mod)   -- (Input_mod_2);
    \draw[-]       (aid_2_mod)   -- (aid_4_mod);
    \draw[-]        (aid_4_mod) -- (Diram0);
    \draw[-]       (aid_3_mod)   -- (aid_5_mod);
    \draw[-]       (aid_5_mod)   -- (Diram3);
    \draw[->]       (Diram3)   -- node[xshift=.25cm, yshift=.25cm]{+}(Sum2);
    \draw[-]       (Diram3)   -- node[xshift=.25cm, yshift=.25cm]{$\mathbf{x}$}(aid_6_mod);
    \draw[-]       (Diram2)   -- (aid_6_mod);

    \draw[->]       (Model)   -- node[xshift=.15cm ,yshift=-.25cm]{-}(Sum2);
    \draw[->]       (Sum2)   -- node[yshift=.25cm]{$\mathbf{w}$}(Map);

     \draw[brace](aid_3_map)--(aid_5_map) node[midway, below]{$ \mathbfcal{M}(\theta)$};
    
 \end{tikzpicture}}\vspace{-.2cm}
\caption{Scheme of the adopted IMC architecture (see \cite{furieri2024learning}).}\label{fig:Control_Scheme}
\end{figure}
\begin{theorem}[Adapted from \text{\cite[Thm 1]{furieri2024learning}}]\label{thm:reformulation}
    Let Assumption~\ref{assumption:L_p_map} hold and consider \eqref{eq:operator_form} with the input sequence chosen as
    \begin{equation}\label{eq:input_sequence_thm}
        \mathbf{u}=\mathbfcal{M}(\mathbf{x}-\mathbf{F}(\mathbf{x},\mathbf{u})),
    \end{equation}
    for a causal operator $\mathbfcal{M}: \ell^{n} \rightarrow \ell^{m}$. Let $\mathbf{K}$ be the operator such that $\mathbf{u}=\mathbf{K}(\mathbf{x})$ is equivalent to \eqref{eq:input_sequence_thm}. \\
    Then, \\
    $(i)$ if $\mathbfcal{M} \in \mathcal{L}_{p}$, then the closed-loop system is $\ell_p$-stable, 
    \\
    $(ii)$ if there is a causal policy $\mathbf{C}$ such that $(\boldsymbol{\Phi}^{\mathbf{x}}[\mathbf{F},\mathbf{C}], \boldsymbol{\Phi}^{\mathbf{u}}[\mathbf{F},\mathbf{C}]) \in \mathcal{L}_{p}$, then
    \begin{equation}
        \mathbfcal{M}=\boldsymbol{\Phi}^{\mathbf{u}}[\mathbf{F},\mathbf{C}],
    \end{equation}
    implies $\mathbf{K}=\mathbf{C}$. \hfill $\blacksquare$
\end{theorem}
Theorem~\ref{thm:reformulation} provides both sufficient and necessary conditions for $\ell_p$-stability. In particular, any controller associated with an $\ell_p$-stable operator $\mathbfcal{M}$ guarantees closed-loop $\ell_p$-stability, and, conversely, any state-feedback law preserving $\ell_p$-stability can be represented through such an operator. According to Theorem~\ref{thm:reformulation}, \eqref{eq:input_dynamics} 
in \eqref{eq:NOC1} can be replaced with 
\begin{equation}
    u_{t}=\mathcal{M}_{t}(w_{[0,t]}),~~~\forall t \in [0,T],
\end{equation}
and $\min_{\mathbf{K}(\cdot)}$ in~\eqref{eq:NOC1_cost} with $\min_{\mathbfcal{M}\in\mathcal{L}_p}$, thus searching in the space of $\mathbfcal{M} \in \mathcal{L}_p$ rather than the set of controllers $\mathbf{K}(\cdot)$ complying with \eqref{eq:NOC1_stability}. Since $\mathcal{L}_p$ is convex and closed under composition, this reformulation allows us to use existing methods to parametrize the operator $\mathbfcal{M}$ as a function of a set of free parameters $\theta \in \mathbb{R}^{d}$, i.e., to define
\begin{equation}\label{eq:parameterization}
    \mathbfcal{M}(\theta) \in \mathcal{L}_p,~~\forall \theta \in \mathbb{R}^{d},
\end{equation}
such that 
\begin{equation}\label{eq:input_parametrized}
    u_{t}=\mathcal{M}_{t}(w_{[0,t]};\theta),~~~\forall t \in [0,T].
\end{equation}
 We stress that the stability guarantee given by Theorem \ref{thm:reformulation} is independent of the convergence of the training algorithm used to train parameters $\theta$. At the same time, such a stability guarantee does not imply the feasibility of the state and input constraints or the optimality of the achieved performance.
 
Leveraging \eqref{eq:input_parametrized}, we derive the following design problem 
\begin{subequations}\label{eq:NOC2}
    \begin{align}
        & \underset{\theta}{\mathrm{minimize}}~~~\mathbb{E}_{w_{[0,T]}}[L(x_{[0,T]},u_{[0,T]})]\\
        &\qquad~~ \mbox{s.t. }~~x_{t}\!=\!\!f_{t}(x_{[0,t-1]},u_{[0,t-1]})\!+\!w_t,\forall t \!\in\! [1,T],\\
        &\qquad\qquad~~~ w_0=x_0,\\
        &\qquad\qquad~~~ u_{t}=\mathcal{M}_{t}(w_{[0,t]};\theta), ~~~~\forall t \!\in\! [0,T], \label{eq:input_dynamicsNOC2}\\
        &\qquad\qquad~~~ x_{t} \!\in\! \mathcal{X},~~u_{t} \!\in\! \mathcal{U}, ~~~~\forall t \!\in\! [0,T],
    \end{align}
\end{subequations} 
In practice, as argued in~\cite[Section 5.B]{furieri2024learning}, directly parameterizing a generic operator in $\mathcal{L}_p$ is challenging, since $\mathcal{L}_p$ is infinite-dimensional. One therefore restricts $\mathbfcal{M}(\theta)$ to a suitable finite-dimensional, stability-preserving subclass, such as Recurrent Equilibrium Networks (RENs)~\citep{revay2023recurrent}, Structured State-Space Models (SSMs)~\citep{orvieto2023resurrecting,massai2025free}, or neural architectures with port-Hamiltonian structure~\citep{zakwanNeuralDistributedControllers2024a}. This restriction to a finite-dimensional, stability-preserving subclass makes the search over stability-preserving controllers tractable, at the price of an approximation error. In particular, the approximation preserves the sufficiency part of the characterization, meaning that closed-loop stability is guaranteed for any value of the parameters $\theta$. Nonetheless, the necessity part may be lost, since the adopted finite-dimensional class may no longer represent all stability-preserving controllers. This may, in turn, introduce an optimality gap with respect to the original infinite-dimensional problem. At the same time, although this restriction may exclude some stability-preserving controllers, the aforementioned classes remain highly expressive. In particular, the RENs used in this work are universal approximators of fading-memory and contracting nonlinear systems~\citep{revay2023recurrent}. Hence, with a sufficiently rich parametrization, the induced approximation error can be expected to be small in practice. 

\section{ADMM-PB: ADMM for Constrained Performance Boosting}\label{Sec:Approach}
Before discussing how to tackle performance boosting with state and input constraints, let us rewrite \eqref{eq:NOC2} in a computationally tractable form. Specifically, we replace the expected value in the loss with the empirical average over $S\geq 1$ scenarios constructed by drawing $w_{[0,T]}^{s}$ samples from the known distribution $\mathcal{D}_{[0,T]}$. Thus, we obtain the following NOC
\begin{subequations}\label{eq:NOC3}
    \begin{align}
        & \underset{\theta}{\mathrm{minimize}}~~~\frac{1}{S}\sum_{s=1}^{S}L(x_{[0,T]}^{s},u_{[0,T]}^{s})\label{eq:NOC3cost}\\
        &\qquad~~ \mbox{s.t. }~~x_{t}^{s}\!=\!\!f_{t}(x_{[0,t-1]}^{s},u_{[0,t-1]}^{s})\!+\!w_t^{s},\forall t \!\in\! [1,T],\label{eq:NOC3_dynamics}\\
        &\qquad\qquad~~~ w_0^{s}=x_0^{s},~~\forall s\! \in [1,S],\\
        &\qquad\qquad~~~ u_{t}^{s}\!=\!\mathcal{M}_{t}(w_{[0,t]}^{s};\theta), \forall t \!\in\! [0,T],\forall s\! \in\! [1,S], \label{eq:input_dynamicsNOC3}\\
        &\qquad\qquad~~~ x_{t}^{s} \!\in\! \mathcal{X},~u_{t}^{s} \!\in\! \mathcal{U}, ~\forall t \!\in\! [0,T],\forall s\! \in [1,S], \label{eq:NOC3_constraints}
    \end{align}
where $x_t^{s} \in \mathbb{R}^{n}$ and $u_t^{s} \in \mathbb{R}^{m}$ are the state and input associated with the sampled process noise $w_{t}^{s}$, for all $t \in [0,T]$ and $s \in [1,S]$. While this reformulation allows us to overcome tractability issues linked with the expectation in the loss of \eqref{eq:NOC2}, it still does not allow us to fully leverage the framework proposed in \citep{furieri2024learning}. Indeed, due to the hard constraints in \eqref{eq:NOC3_constraints}, solving \eqref{eq:NOC3} with Gradient Descent (GD) (see \cite{ruder2016overview} for an overview) as proposed therein could become computationally prohibitive \citep{imposinghardconstraintsdeep}.
\end{subequations}

To retain the main features (and advantages) of the approach in \citep{furieri2024learning}, while not altering the structure of the control law (e.g., by including projection layers in a neural controller as proposed in \citep{paulson2020approximate,grontas2025pinet}) not to undermine closed-loop stability, we propose to optimize the free parameters $\theta$ of \eqref{eq:parameterization} by using the Alternating Direction Method of Multipliers (ADMM, \cite{boyd2011distributed}).  

\subsection{ADMM reformulation of performance boosting}
To unveil the structure exploited by ADMM, let us make explicit the dependence of the sampled state and input trajectories on the controller parameters \(\theta\). For each scenario \(s\in[1,S]\), let us thus denote the state and input trajectories generated by \eqref{eq:NOC3_dynamics}--\eqref{eq:input_dynamicsNOC3} for the sampled disturbance sequence \(w^{s}_{[0,T]}\) as
\begin{equation}
    X^{s}(\theta)=x^{s}_{[0,T]},
\qquad
U^{s}(\theta)=u^{s}_{[0,T]},~s=1,\ldots,S. 
\end{equation}
Accordingly, we can equivalently rewrite the objective in \eqref{eq:NOC3cost} as  
\begin{equation}
    J(\theta)=\frac{1}{S}\sum_{s=1}^{S} L(X^{s}(\theta),U^{s}(\theta)),
\end{equation}
leading to the following equivalent, yet more compact, formulation of \eqref{eq:NOC3}:
\begin{subequations}\label{eq:NOC4_new}
\begin{align}
    &\underset{\theta}{\mathrm{minimize}}
    && \frac{1}{S}\sum_{s=1}^{S} L(X^{s}(\theta),U^{s}(\theta)) \\
    &\mathrm{subject\;to}
    && x_t^s(\theta)\in\mathcal{X},\quad
      u_t^s(\theta)\in\mathcal{U},
      \qquad \forall t\in[0,T],\;\forall s\in[1,S].
\end{align}
\end{subequations}
In addition, by stacking all scenario-wise trajectories as 
\begin{equation}\label{eq:trajectories_compact}
    X(\theta)= \begin{bmatrix}
        X^{1}(\theta)\\
        \vdots\\
        X^{S}(\theta)
    \end{bmatrix},
\qquad
U(\theta)= \begin{bmatrix}
        U^{1}(\theta)\\
        \vdots\\
        U^{S}(\theta)
    \end{bmatrix}.
\end{equation}
and introducing the constraint set
\begin{equation}\label{eq:compact_constraints}
\mathcal C = \Big\{(X,U)~\mathrm{s.t.}~ X\in\mathcal{X}^{S(T+1)},\; U\in\mathcal{U}^{S(T+1)}\Big\},
\end{equation}
we can further simplify \eqref{eq:NOC4_new}, equivalently recasting it as
\begin{subequations}\label{eq:NOC_compact}
    \begin{align}
    &\underset{\theta}{\mathrm{minimize}} &&J(\theta)\label{eq:NOC_compact_J}\\
    &\mathrm{subject\;to} &&
    \big(X(\theta),U(\theta)\big)\in\mathcal C.\label{eq:NOC_compact_C}
\end{align}
\end{subequations}
Finding a solution for \eqref{eq:NOC_compact} still retains the same challenges as the ones of \eqref{eq:NOC3}, since constraints act on trajectories that depend nonlinearly on $\theta$ through the closed-loop rollout. To decouple the trajectory generation from the constraint enforcement step, we introduce copy variables for the trajectories of each scenario, i.e., $X_p^{s}=x_{[0,T]}^{s,p}$ and $U_p^{s}=u_{[0,T]}^{s,p}$ for all $s \in [1,S]$. We then stack them into
\begin{equation*}
   X_p=\begin{bmatrix}
       X_p^{1}\\
       \vdots\\
       X_p^{S}
   \end{bmatrix},
\qquad
U_p=\begin{bmatrix}
       U_p^{1}\\
       \vdots\\
       U_p^{S}
   \end{bmatrix}, 
\end{equation*}
and rewrite \eqref{eq:NOC_compact_J}-\eqref{eq:NOC_compact_C} as
\begin{subequations}\label{eq:NOC_ADMM_new}
\begin{align}\label{eq:two_block_ADMM}
    &\underset{\theta,X_p,U_p}{\mathrm{minimize}}
    && J(\theta)+\mathcal{I}_{\mathcal C}(X_p,U_p)\\
    &\mathrm{subject\;to}
    && X(\theta)=X_p,\qquad U(\theta)=U_p, \label{eq:copy_variables}
\end{align}
\end{subequations}
where $\mathcal{I}_{\mathcal C}$ denotes the indicator function of $\mathcal C$ in \eqref{eq:compact_constraints}. This formulation (which is equivalent to \eqref{eq:NOC3}) has revealed the two-block structure and the nonlinear coupling in our optimization problem. The first block is the controller parameter $\theta$, which determines the rollout $(X(\theta),U(\theta))$, while the second one is the copy pair $(X_p,U_p)$, on which the hard state and input constraints are imposed. As already mentioned, ADMM is therefore attractive in our context because it separates $(i)$ a rollout-based optimization step in $\theta$, retaining the same structure as unconstrained PB, from $(ii)$ a constraint-enforcement step on $(X_p,U_p)$, which reduces to Euclidean projections onto the convex sets $\mathcal X$ and $\mathcal U$.

Towards defining the ADMM update procedure, let us then introduce $\Lambda$, which stacks the scaled dual variables as
\begin{equation*}
    \Lambda=\{\lambda_{[0,T]}^{x,s},\lambda_{[0,T]}^{u,s}\}_{s=1}^{S},    
\end{equation*}
and let \(\rho>0\) be a tunable penalty parameter. These elements allow us to write the augmented Lagrangian associated with \eqref{eq:NOC_ADMM_new} as
\begin{equation}\label{eq:augmented_lagrangian_new}
    \mathscr L(\theta,X_p,U_p,\Lambda)
    =
    J(\theta)+\mathcal{I}_{\mathcal C}(X_p,U_p)
    +\frac{\rho}{2}L^{\mathrm a}(\theta,X_p,U_p,\Lambda),
\end{equation}
with
    \begin{align}\label{eq:augmenting_term}
    L^{\mathrm a}(\theta,X_p,U_p,\Lambda)
    =
    \frac{1}{S}\sum_{s=1}^{S}\Big(\|x^{s}_{[0,T]}(\theta)-x^{s,p}_{[0,T]}+\lambda^{x,s}_{[0,T]}\|_2^2+\|u^{s}_{[0,T]}(\theta)-u^{s,p}_{[0,T]}+\lambda^{u,s}_{[0,T]}\|_2^2
    \Big).
    \end{align}    
Based on this augmented Lagrangian, 
the ADMM-based scheme for constrained Performance Boosting consists of the following iterations
\begin{subequations}\label{eq:ADMM_PB}
\begin{align}
    &\theta^{(j+1)}
    \leftarrow
    \arg\min_{\theta}\;
    \mathscr L\big(\theta,X_p^{(j)},U_p^{(j)},\Lambda^{(j)}\big),
    \label{eq:ADMM_PB_step1}\\
    &(X_p^{(j+1)},U_p^{(j+1)})
    \leftarrow
    \arg\min_{X_p,U_p}\;
    \mathscr L\big(\theta^{(j+1)},X_p,U_p,\Lambda^{(j)}\big),
    \label{eq:ADMM_PB_step2}\\
    &\lambda_{[0,T]}^{x,s,(j+1)}
    \leftarrow
    \lambda_{[0,T]}^{x,s,(j)}
    +x^{s}_{[0,T]}(\theta^{(j+1)})
    -x_{[0,T]}^{s,p,(j+1)},
    \qquad \forall s\in[1,S],
    \label{eq:ADMM_PB_step3_1}\\
    &\lambda_{[0,T]}^{u,s,(j+1)}
    \leftarrow
    \lambda_{[0,T]}^{u,s,(j)}
    +u^{s}_{[0,T]}(\theta^{(j+1)})
    -u_{[0,T]}^{s,p,(j+1)},
    \qquad \forall s\in[1,S].
    \label{eq:ADMM_PB_step3_2}
\end{align}
\end{subequations}
As discussed next, the key point of this procedure is that \eqref{eq:ADMM_PB_step1} updates the controller parameters through the same rollout-based optimization structure as in unconstrained PB, whereas \eqref{eq:ADMM_PB_step2} enforces feasibility only for the auxiliary copied trajectories, via their projections onto $\mathcal X$ and $\mathcal U$. Meanwhile, the feasibility of the rollout trajectories generated by $\theta$ is recovered only when the primal residuals between rollout and copy variables are driven to zero. Such a condition is nonetheless likely not achieved if ADMM is terminated early or if it does not converge. 

\begin{remark}[Ordering of ADMM-PB's steps]\label{remark:ordering}
    The order in which the steps in \eqref{eq:ADMM_PB} are executed can be modified by changing how ADMM-PB is initialized. By inverting the order of the first and second steps in \eqref{eq:ADMM_PB} one can take advantage of the approach proposed in \citep{furieri2024learning} to warm-start $\theta$. \textcolor{black}{Instead, when maintaining the order in \eqref{eq:ADMM_PB}, one can solve the performance boosting problem of \citep{furieri2024learning} and then project the resulting state and input sequences onto the constraint sets to initialize $X_{p}$ and $U_{p}$.} 
\end{remark}
\begin{remark}[Stability \& early stopping]
\label{rem:early_stopping}
Thanks to Theorem~\ref{thm:reformulation} and the parameterization of $\mathbfcal{M}$ in~\eqref{eq:parameterization}, the controller obtained at any ADMM-PB iteration remains within the chosen stable parametrized class. Indeed, for every value of the parameters $\theta$, the induced operator $\mathbfcal{M}(\theta)$ still belongs to $\mathcal{L}_p$. In turn, by  Theorem~\ref{thm:reformulation}, this property is sufficient to make the closed-loop system $\ell_p$-stable in the sense of Definition~\ref{def:closed_loop_stab}. Therefore, this stability guarantee is independent of the convergence of ADMM-PB and of the adopted termination criterion. If ADMM-PB is stopped early, the ADMM residuals and, hence, the achieved performance and the constraint violations depend on the stopping point (even if the resulting controller preserves the built-in stability guarantee). Thus, stability should not be linked to ADMM convergence, which instead drives performance optimality and handles constraints during training.

\end{remark}

\begin{remark}[ADMM-PB and optimization-based constrained control schemes]
    Unlike approaches such as Model Predictive Control (MPC), where the closed-loop guarantees are typically tied to the successful solution of an online or embedded optimization problem, our setting relies on a different mechanism. In particular, since $\mathbfcal{M}(\theta)\in\mathcal{L}_p$ for every $\theta$ by construction, 
    $\ell_p$-stability is preserved according to Theorem~\ref{thm:reformulation} irrespectively of the iteration at which the ADMM-PB procedure in \eqref{eq:ADMM_PB} is stopped. In the approach we propose, ADMM is therefore used as an offline training mechanism to optimize an empirical performance objective and promote constraint satisfaction, without affecting the stability guarantee provided by the controller parametrization. 
\end{remark}

\subsection{Augmented performance boosting}
Let us first focus on the ADMM-PB step in~\eqref{eq:ADMM_PB_step1}, namely
\begin{equation}\label{eq:ADMM_step1_PB}
    \theta^{(j+1)}
    \leftarrow
    \arg\min_{\theta}\;
    J(\theta)
    +\frac{\rho}{2}L^{\mathrm a}\!\left(\theta,X_p^{(j)},U_p^{(j)},\Lambda^{(j)}\right).
\end{equation}
This subproblem retains the same rollout-based structure as the sampled performance-boosting problem in~\eqref{eq:NOC3}. Indeed, the decision variable is still $\theta$, i.e., the NN controller's parameters, and the objective function is evaluated along the closed-loop trajectories induced by them. The main difference is that the hard state and input constraints in~\eqref{eq:NOC3_constraints} are no longer enforced explicitly. Instead, they are implicitly accounted for through the quadratic term $L^{\mathrm a}\!\left(\theta,X_p^{(j)},U_p^{(j)},\Lambda^{(j)}\right)$ defined in~\eqref{eq:augmenting_term}, which penalizes the mismatch between the trajectories generated by $\theta$ and their constrained copies, promoting feasibility. 
We can thus exploit the same training strategy as in~\cite{furieri2024learning}. In particular, since the closed-loop rollout is already implicit in the definitions of $J(\theta)$, $X(\theta)$, and $U(\theta)$, problem~\eqref{eq:ADMM_step1_PB} can be addressed directly via gradient-based methods (see~\cite{ruder2016overview} for an overview), i.e.,
\begin{equation}\label{eq:gradient_descent}
    \theta^{(i+1)}\!=\!\theta^{(i)\!}+\!\eta\nabla_{\theta} \!\!\left[J(\theta)\!+\!\frac{\rho}{2}L^{\mathrm{a}}(\theta,X_p^{(j)},U_p^{(j)},\Lambda^{(j)})\right]\!\!\bigg|_{\theta^{(i)}},
\end{equation}
where $i=0,1,\ldots,E-1$ denotes the inner gradient iteration and $\eta\geq 0$ is a tunable learning rate. Therefore, the $\theta$-subproblem is approximately solved by performing $E$ gradient-based epochs/iterations of the update in \eqref{eq:gradient_descent}. 
Note that, solving \eqref{eq:ADMM_step1_PB} via gradient descent implies almost sure (thus, asymptotically in the number of gradient iterations) convergence to a stationary point of \eqref{eq:ADMM_step1_PB} (see \cite{lee2016gradient}).

\subsection{Projection onto the constraint sets}
We now consider the second ADMM step in~\eqref{eq:ADMM_PB_step2}. For fixed $\theta^{(j+1)}$ and $\Lambda^{(j)}$, the corresponding optimization problem is
\begin{equation}\label{eq:ADMM_PB_step2_PB}
    \underset{X_p,U_p}{\mathrm{minimize}}\;
    \mathcal{I}_{\mathcal C}(X_p,U_p)
    +\frac{\rho}{2}L^{\mathrm a}\big(\theta^{(j+1)},X_p,U_p,\Lambda^{(j)}\big).
\end{equation}
Since \(\mathcal{I}_{\mathcal C}\) is the indicator function of the constraint set \(\mathcal C\), problem~\eqref{eq:ADMM_PB_step2_PB} is equivalent to
\begin{equation}\label{eq:ADMM_PB_step2_proj}
    \underset{(X_p,U_p)\in\mathcal C}{\mathrm{minimize}}\;
    L^{\mathrm a}\big(\theta^{(j+1)},X_p,U_p,\Lambda^{(j)}\big).
\end{equation}
Hence, the copy update amounts to finding the feasible trajectories $(X_p,U_p)$ closest, in Euclidean norm, to the shifted trajectories resulting from the previous step, thus $\big(X(\theta^{(j+1)}),U(\theta^{(j+1)})\big)+\Lambda^{(j)}$. Since $\mathcal C$ is convex and separable with respect to states and inputs for the different scenarios, the solution is given by Euclidean projections onto the sets $\mathcal{X}^{T+1}$ and $\mathcal{U}^{T+1}$~\citep{boyd2011distributed}, namely
\begin{subequations}\label{eq:projections}
\begin{align}
    x_{[0,T]}^{s,p,(j+1)}
    &\leftarrow
    \Pi_{\mathcal{X}^{T+1}}
    \Big(
    x_{[0,T]}^{s}(\theta^{(j+1)})
    +\lambda_{[0,T]}^{x,s,(j)}
    \Big),
    \qquad \forall s\in[1,S],\\
    u_{[0,T]}^{s,p,(j+1)}
    &\leftarrow
    \Pi_{\mathcal{U}^{T+1}}
    \Big(
    u_{[0,T]}^{s}(\theta^{(j+1)})
    +\lambda_{[0,T]}^{u,s,(j)}
    \Big),
    \qquad \forall s\in[1,S].
\end{align}
\end{subequations}

\begin{remark}[Copy variables \& their shortcomings]
As an alternative to the auxiliary variables introduced in \eqref{eq:copy_variables}, one could consider copying the free parameters \(\theta\), similarly to what is proposed in~\cite{pauli2021training}. While this choice would reduce the number of optimization variables when \(d< S(T+1)(m+n)\) and would yield a linearly coupled reformulation, it would shift the hard constraints to the parameter space. Consequently, the projection step would no longer amount to Euclidean projections onto the convex sets, but to a projection onto a generally nonconvex set induced by the nonlinear rollout map. In turn, this shift increases the complexity of the projection stage or requires to approximate it (see, e.g.,\cite{Diamond02018}).
\end{remark}

\subsection{Positioning with respect to theoretical results on ADMM convergence}
\label{subsec:convergence_discussion}
By inheriting the approach proposed in~\cite{furieri2024learning} to solve \eqref{eq:ADMM_PB_step1}, the proposed ADMM-PB scheme is an \emph{inexact nonconvex ADMM-type procedure}. Indeed, while the projection step admits a closed-form solution\footnote{As already mentioned, it reduces to Euclidean projections onto the convex sets of admissible state and input trajectories}, the optimization over the NN parameters $\theta$ is a nonconvex training problem that we solve only approximately via a finite number of gradient-based iterations. For this reason, classical convergence results for ADMM do not directly apply, as they are limited to two-block convex problems with linear coupling constraints~\citep{boyd2011distributed}. While more recent nonconvex ADMM analyses significantly extend convergence further to this setting, they still require structural assumptions that do not directly match the present formulation. Examples range from special lifted architectures in neural-network training or specific treatments of nonlinear equality constraints~\citep{bai2025inexact,frangella2023linear,wang2019global}. Indeed, in the problem we get after variable splitting (i.e., \eqref{eq:NOC_ADMM_new}), the two ADMM blocks are coupled through the nonlinear mapping of the trajectories stacked in $X(\theta)$ and $U(\theta)$ generated by the closed-loop rollout. Such nonlinearity is induced by the system dynamics and by the chosen, stable-by-design controller parameterization.


Among recent findings, the convergence result that more closely matches our setting is the one proposed in~\cite{el2025convergence}. Specifically, this work analyzes the convergence of two-block, non-smooth nonconvex problems with equality constraints, with the latter being nonlinear in one block and linear in the other one. While our reformulation structurally matches this template, as the equality constraints in \eqref{eq:NOC_ADMM_new} are nonlinear in $\theta$ and linear in the auxiliary variables (that are also easy to project), the convergence guarantees in~\cite{el2025convergence} are established for a different algorithmic scheme. Indeed, differently from the approach inherited from \cite{furieri2024learning}, the strategy proposed in \cite{el2025convergence} relies on a linearization of the nonlinear coupling 
inside the augmented Lagrangian, introduces an additional dynamic quadratic regularization, and controls the inexactness of the nonlinear block through explicit descent and stationary conditions. In addition, the analysis performed in~\cite{el2025convergence} relies on assumptions such as a bounded decrease of the ADMM-subproblems' losses along the solver iterations and further regularity properties of the constrained block. However, verifying them in a rollout-based setting like the one considered in ADMM-PB is nontrivial.

Convergence analyses for ADMM-based approaches tailored to neural-network training also do not directly cover our case. In fact, existing results are typically derived for specific lifted formulations of feedforward or residual architectures~\citep{wang2019admm,zeng2021admm}. Instead, in our setting the optimization variable is the controller parameter \(\theta\), and the coupling is induced by a nonlinear control-oriented rollout rather than by a layer-wise factorization. Instances that show encouraging practical performance of empirical ADMM-based training methods for the same stable-by-design neural architecture of our controller~\citep{taylor2016training,guan2021pdladmm} also do not provide any convergence proof. 

Developing a dedicated convergence theory for this rollout-based, stable-by-design ADMM scheme is therefore left for future work. Meanwhile, the goal of this paper is to show that the proposed splitting is practically effective for constrained performance boosting while preserving the original controller architecture, laying the foundations for further theoretical investigations.

\section{Practical aspects of ADMM-PB}\label{sec:practical_aspects}
\begin{algorithm}[!tb]
	\caption{ADMM-PB}
	\label{algo:1}
	\textbf{Input}: ADMM and gradient descent step sizes $\rho,\eta \geq 0$; optimization variables initialization $X_p^{(0)}$, $U_p^{(0)}$, $\Lambda^{(0)}$.
	\vspace*{.1cm}\hrule\vspace*{.1cm}
	\begin{enumerate}[label=\arabic*., ref=\theenumi{}]
    \item \textbf{for} $j=0,1,\ldots$ \textbf{do}
    \begin{enumerate}[label=\theenumi{}.\arabic*., ref=\theenumi{}.\theenumii{}]
		\item \textbf{Solve} \eqref{eq:ADMM_step1_PB} to update the free parameters with gradient descent (see \eqref{eq:gradient_descent});
        \item \textbf{Project} variables onto the desired constraint sets as in \eqref{eq:projections};
        \item \textbf{Update} the Lagrange multipliers following \eqref{eq:ADMM_PB_step3_1}-\eqref{eq:ADMM_PB_step3_2};
        \end{enumerate}
        \item \textbf{until} a pre-defined stopping criterion is satisfied.
	\end{enumerate}
	\vspace*{.1cm}\hrule\vspace*{.1cm}
	~~\textbf{Output}: Parameters $\theta$ of the map in \eqref{eq:parameterization}.
\end{algorithm} 
The main steps of ADMM-PB are summarized in Algorithm~\ref{algo:1}, which also highlights some of the practical choices, beyond initialization, that users have to make to run it (see Remark~\ref{remark:ordering} for a discussion on it). Apart from selecting the structure of the parameterization in \eqref{eq:parameterization}, users are indeed required to choose a termination criterion, as well as the step sizes $\rho$ and $\eta$. In this section, we discuss possible practical strategies to automate the choice of the former and to adapt step-sizes along the ADMM-PB iterations. We also briefly discuss the computational complexity of the resulting procedure.

\subsection{Termination criteria}\label{subsec:termination_criteria}
We monitor practical indicators for termination criteria inspired by the residual-based strategy of~\cite[Section~3.3]{boyd2011distributed}. Nonetheless, since ADMM-PB features nonlinear coupling through the closed-loop trajectories together with an inexact $\theta$-update, we stress that this standard primal--dual stopping criteria do not 
admit the same convergence interpretation they have in the classical linearly coupled ADMM. Hence, in our setting, these termination criteria have to be intended as a heuristic.

To decide when to stop the ADMM-PB iterations, we track the mismatch between primal and copy variables and the variation of the projected variables across successive iterations. In particular, at the 
$j$-th ADMM-PB iteration 
we check
\begin{subequations}
\begin{equation}\label{eq:residuals}
    r^{(j)}=\begin{bmatrix}
        r^{x,(j)}\\
        r^{u,(j)}
    \end{bmatrix},~~~\delta^{(j)}=-\rho\begin{bmatrix}
        r^{x,p,(j)}\\
        r^{u,p,(j)}
    \end{bmatrix}\!,
\end{equation} 
where
\begin{equation*}
    r^{\xi,(j)}\!=\! \begin{bmatrix}
        \xi^{1,(j)}_{[0,T]}-\xi^{1,p,(j)}_{[0,T]}\\
        \vdots\\
        \xi^{S,(j)}_{[0,T]}-\xi^{S,p,(j)}_{[0,T]}
    \end{bmatrix}\!,  \qquad 
        r^{\xi,p,(j)}\!=\! \begin{bmatrix}
        \xi^{1,p,(j)}_{[0,T]}-\xi^{1,p,(j-1)}_{[0,T]}\\
        \vdots\\
        \xi^{S,p,(j)}_{[0,T]}-\xi^{S,p,(j-1)}_{[0,T]}
    \end{bmatrix}\!,
\end{equation*}
and $\xi$ denotes a placeholder either for $x$ or $u$. 
\end{subequations}
Here, \(r^{(j)}\) measures the mismatch between primal and copy variables, while \(\delta^{(j)}\) tracks the variation of the copy variables across ADMM iterations.

Inspired by the criterion in~\citep{boyd2011distributed}, ADMM-PB iterations are heuristically terminated when both indicators are sufficiently small, i.e.,
\begin{equation}\label{eq:termination}
    \|r^{(j)}\|_{2}\leq \epsilon^{r,(j)},~~ \|\delta^{(j)}\|_{2}\leq \epsilon^{\delta,(j)},
\end{equation}
where $\epsilon^{r,(j)},\epsilon^{\delta,(j)}\geq 0$ are tolerances that could be either predefined (i.e., $\epsilon^{r,(j)}=\epsilon^{r}$ and $\epsilon^{\delta,(j)}=\epsilon^{\delta}$ for all $j=0,1,\ldots$) or adjusted iteratively along the ADMM iterations. In the second case, the primal and dual tolerances can be iteratively modified utilizing the vectors stacking the state and input sequences and the corresponding copy variables at the $j$-th ADMM-PB iteration over the $S$ scenarios, namely
\begin{align*}
    z_{[0,T]}^{(j)}=\begin{bmatrix}
       z_{[0,T]}^{1,(j)}\\
       \vdots\\
       z_{[0,T]}^{S,(j)}
    \end{bmatrix},~~~z_{[0,T]}^{p,(j)}=\begin{bmatrix}
       z_{[0,T]}^{1,p,(j)}\\
       \vdots\\
       z_{[0,T]}^{S,p,(j)}
    \end{bmatrix}
\end{align*}
where 
\begin{equation*}
    z_{[0,T]}^{s,(j)}=\mathrm{col}(x_{[0,T]}^{s,(j)},u_{[0,T]}^{s,(j)}),~~z_{[0,T]}^{s,p,(j)}=\mathrm{col}(x_{[0,T]}^{s,p,(j)},u_{[0,T]}^{s,p,(j)}),
\end{equation*}
for all $s \in [1,S]$. Using these quantities, $\epsilon^{r,(j)}$ and $\epsilon^{\delta,(j)}$ can be adjusted as
\begin{align}
    & \epsilon^{r,(j)}=\sqrt{c}\epsilon^{\mathrm{abs}}+ \epsilon^{\mathrm{rel}}\max\{\|z_{[0,T]}^{(j)}\|_{2},\|z_{[0,T]}^{p,(j)}\|_{2}\},\\
    & \epsilon^{\delta,(j)}=\sqrt{o}\epsilon^{\mathrm{abs}}+ \epsilon^{\mathrm{rel}}\|\Lambda^{(j)}\|_{2},
\end{align}
where $c=S(T+1)(n+m)$ and $o=c+d$ are the number of constraints and optimization variables, respectively, while $\epsilon^{\mathrm{abs}}, \epsilon^{\mathrm{rel}}\geq 0$ control the trade-off between the dimension of the residuals and the magnitude of the variables that ultimately characterize them. These tolerances therefore affect when ADMM-PB is terminated, and hence the level of matching between the rollout trajectories and the projected auxiliary trajectories. As discussed in Remark~\ref{rem:early_stopping}, this dependence should not be confused with a dependence of closed-loop stability on the selected tolerances. Indeed, under our assumptions, stability follows from the chosen (stable) controller parametrization and is preserved irrespective of the adopted termination criterion. In contrast, constraint handling during training depends on the chosen tolerances and residual-dependent termination criteria. In particular, as smaller residuals indicate better agreement between rollout and projected trajectories, exact constraint satisfaction across the sampled training trajectories would require the corresponding feasibility residuals to vanish.


\subsection{Tuning the step sizes in ADMM-PB}\label{sec:tuning}
We now focus on the key hyperparameters of ADMM-PB, namely the step sizes $\rho$ and $\eta$. Before doing so, we remark that the following update rules should be interpreted as practical heuristics inspired by the ADMM literature, rather than as theoretically justified choices for our setting, as ADMM-PB involves nonlinear coupling and an inexact $\theta$-update. 

A practical way to adapt $\rho$ over the iterations is to use the residual-balancing strategy proposed in~\citep{he2000alternating}, which aims at keeping the ratio between the primal and dual stopping indicators within a user-defined factor $\mu\geq 0$. Specifically, over iterations, one can use the following rule
\begin{equation}\label{eq:update_rho}
    \rho^{(j+1)}=\begin{cases}
        \tau^{\mathrm{inc}}\rho^{(j)},~~&\mbox{if } \|r^{(j)}\|_{2}>\mu \|\delta^{(j)}\|_{2},\\
        \tau^{\mathrm{dec}}\rho^{(j)},~~&\mbox{if } \|\delta^{(j)}\|_{2}>\mu \|r^{(j)}\|_{2},\\
        \rho^{(j)},~~&\mbox{otherwise},
    \end{cases}
\end{equation}
with $r^{(j)}$ and $\delta^{(j)}$ defined as in \eqref{eq:residuals}, $\tau^{\mathrm{inc}}>1$ and $\tau^{\mathrm{dec}} \in (0,1)$ being two user-defined parameters, to be selected along with the initial step size $\rho^{(0)}$. Since we employ a scaled version of ADMM, when $\rho$ is updated, we correspondingly rescale the dual variables before the next iteration, i.e.,
\begin{equation}
    \lambda_{[0,T]}^{\xi,s,(j)}\leftarrow \begin{cases}
        \frac{\lambda_{[0,T]}^{\xi,s,(j)}}{\tau^{\mathrm{inc}}},~~&\mbox{if } \|r^{(j)}\|_{2}>\mu \|\delta^{(j)}\|_{2},\\
         \frac{\lambda_{[0,T]}^{\xi,s,(j)}}{\tau^{\mathrm{dec}}},~~&\mbox{if } \|\delta^{(j)}\|_{2}>\mu \|r^{(j)}\|_{2},\\
         \lambda_{[0,T]}^{\xi,s,(j)},~~&\mbox{otherwise},
    \end{cases}
\end{equation}
with $\xi$ being once more a placeholder for $x$ and $u$. 

While $\rho$ governs the outer ADMM iterations, the approximate solution of \eqref{eq:ADMM_PB_step1} is driven by the learning rate $\eta$ in \eqref{eq:gradient_descent}, which strongly affects the quality of the rollout-based parameter update.
In our implementation, rather than adapting $\eta$ within the inner gradient iterations, we update it across outer ADMM iterations according to
\begin{equation}\label{eq:step_size_choice}
    \eta^{(j+1)}=\eta^{(0)}\gamma^{\lfloor \frac{(j)}{\mathcal{J}} \rfloor},
\end{equation}
where $\eta^{(0)}$ is the initial learning rate\textcolor{black}{, and $\gamma \in (0,1)$ dictates the decay of the step size taking place every $\mathcal{J}\geq 1$ iterations}. 
This choice is meant to progressively reduce the aggressiveness of the $\theta$-updates as ADMM-PB proceeds, so as to favor more stable warm-started solutions of the $\theta$-subproblem and, empirically, a better balance between performance improvement and feasibility. In practice, as the projected variables and rollout trajectories tend to get closer, decreasing $\eta$ often leads to smaller variations in the copy variables and hence to a more stable evolution of the stopping indicators. This behavior motivates the combined use of \eqref{eq:update_rho} and \eqref{eq:step_size_choice}, although we do not claim a formal guarantee for this tuning strategy. To avoid excessive reductions of the gradient step size, we impose a lower bound $\bar{\eta}$, i.e., $\eta^{(j+1)}$ is given by \eqref{eq:step_size_choice} as long as $\eta^{(j+1)}\geq \bar{\eta}$, and otherwise we set $\eta^{(j+1)}=\bar{\eta}$.

\subsection{Computational complexity}
We now discuss the computational cost of Algorithm~\ref{algo:1}, focusing on a single ADMM-PB iteration. To this end, let $E$ denote the number of gradient-based inner iterations used to approximately solve the $\theta$-subproblem in~\eqref{eq:ADMM_step1_PB}. Moreover, let $C_{\mathrm{rb}}$ denote the cost of one forward/backward closed-loop rollout over one time step for one scenario. Note that this quantity depends on the complexity of the system to be controlled and on the adopted stable parameterization for $\mathbfcal{M}(\theta)$. Then, for \(S\) sampled scenarios and horizon \(T\), the cost of one outer ADMM iteration can be expressed as
\begin{equation}\label{eq:complexity_admm_pb}
    \mathcal O\!\left(
    ES(T+1)C_{\mathrm{rb}}
    +
    S(T+1)C_{\Pi}
    +
    S(T+1)(n+m)
    \right),
\end{equation}
where $C_{\Pi}$ denotes the cost of projecting one state-input pair onto the admissible sets $\mathcal{X}$ and $\mathcal{U}$, while the last term accounts for the dual-variable update. The first term in~\eqref{eq:complexity_admm_pb}, corresponding to the rollout-based $\theta$-update, represents the dominant part of the training complexity in practice. Importantly, this step has the same structure as the original unconstrained PB training problem, since it still consists of repeated closed-loop rollouts and gradient-based parameter updates. The second and third terms arise from the copy-variable projection and the dual update. In particular, when $\mathcal{X}$ and $\mathcal{U}$ are simple convex sets, such as boxes or Cartesian products of low-dimensional convex sets, the projection step is separable across time and scenarios, and $C_{\Pi}$ is linear in the state and input dimensions. In this case, \eqref{eq:complexity_admm_pb} reduces to
\begin{equation}\label{eq:complexity_admm_pb_simple}
    \mathcal O\!\left(
    E\,S(T+1)\,C_{\mathrm{rb}}
    +
    2S(T+1)(n+m)
    \right),
\end{equation}
Therefore, when considering simple convex constraints, ADMM-PB has the same rollout-based training complexity as unconstrained PB in~\cite{furieri2024learning}, with the addition of a projection and dual-update step. 
In this scenario, for a fixed controller architecture, the cost per iteration scales linearly with the number of scenarios $S$ and the horizon length $T$. 

By featuring additional projection and dual-update steps, it is worth pointing out that ADMM-PB is generally not cheaper per outer iteration than its unconstrained counterpart in~\cite{furieri2024learning}. This increase in computational complexity is nonetheless due to a more systematic treatment of state and input constraints to promote their feasibility. Indeed, while retaining the original rollout-based PB update, constraints are handled during training through an explicit feasibility step, rather than only through penalty tuning. At the same time, the additional computational complexity still has to be handled in the offline training phase, having no impact at deployment time.

\section{Benchmark examples}
\label{sec:numerical_example}
To evaluate the performance of the proposed approach on two complementary settings, we consider two case studies. The first is a benchmark similar to the one considered in \citep{furieri2024learning,furieri2025mad}, which highlights the main practical advantage of ADMM-PB over penalty-based constrained training, namely the avoidance of delicate penalty tuning. The second is a three-tank system benchmark \citep{massai2025l2ru,massai2024unconstrained}, which serves as a more classical control example and shows how ADMM-PB can enforce constraints starting from a PB controller trained with soft penalties, without modifying the PB structure nor compromising its stability-by-design properties. Together, these benchmarks allow us to isolate the main mechanism of ADMM-PB and to compare constrained and unconstrained training in two representative scenarios\footnote{The code to reproduce our results is available at \url{https://github.com/GiacomelliGianluca/Safe_Performance_Boosting.git}.}.

\subsection{Constrained control of a point-mass robot}
\begin{figure}
\centering
\begin{tabular}{cc}
    \subfigure{\includegraphics[scale=1]{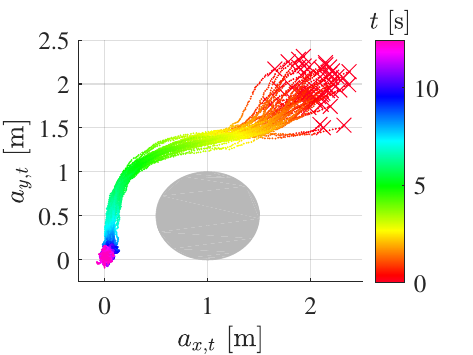}} \hspace*{-.35cm}&\hspace*{-.35cm}  \subfigure{\includegraphics[scale=1]{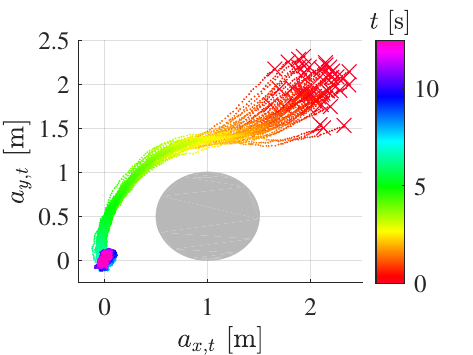}}\\
    (a) ADMM-PB & (b) Baseline for $\omega=10^{2}$    
\end{tabular}
 \caption{ADMM-PB \emph{vs} penalty-based PB baseline: robot position over time (as indicated in the colorbar) with red $\times$ denoting the initial positions. Each trajectory refers to one of the $50$ testing scenarios, showing that overall the baseline with $\omega=10^2$ leads to a faster convergence to the origin, yet trajectories are slightly less consistent across scenarios.}\label{fig:plane_traj}
\end{figure}
\begin{table}[!tb]
    \centering
     \caption{Parameters of the point-mass robot dynamics \eqref{eq:point_mass_sys}.}
    \label{tab:sys_param}
    \begin{tabular}{cccc}
        $M$ [kg] & $\beta_1$ [kg s$^{-1}$] & $\beta_{2}$ [N] & $T_s$ [s] \\
        \hline
         1 & 1 & 0.1 & 0.05\\
         \hline
    \end{tabular}
\end{table}
We consider a single point mass robot that we aim to bring to the origin of the 2D plane while avoiding a circular obstacle on its path centered in $(a_x^{\mathrm{obs}},a_y^{\mathrm{obs}})=(1 ~\mathrm{[m]},0.5~ \mathrm{[m]})$ with radius $0.5$ [m] (see \figurename{~\ref{fig:plane_traj}). The dynamics of the robot are described by
\begin{equation}\label{eq:point_mass_sys}
    x_{t}\!=\!x_{t-1}\!+\!T_s\!\begin{bmatrix}
        q_{t-1}\\
        M^{-1}(\beta_{1}q_{t-1}\!+\!\beta_{2}\tanh{(q_{t-1})}\!+\!F_{t-1})
    \end{bmatrix}\!\!+\!w_t,
\end{equation}
with the parameters reported in \tablename{~\ref{tab:sys_param}}, where $F_t \in \mathbb{R}^{2}$ represents the input fed to the robot, and $x_t \in \mathbb{R}^{4}$ (with a slight abuse of notation) indicates both the state and one of the dimensions of the 2D plane. In our setting, $x_0$ is assumed Gaussian distributed with mean $\smallmat{2 & 2 & \boldsymbol{0}}^{\top}$ and standard deviation $\smallmat{0.2 & 0.2 & \boldsymbol{0}}^{\top}$, while, for $t\geq 1$, $w_t$ in \eqref{eq:point_mass_sys} follows a Gaussian distribution with zero mean and standard deviation $0.005I$. The components of the state $x_t$ are the 2D-positions (in [m]), $a_t \in \mathbb{R}^{2}$, and velocities $q_t \in \mathbb{R}^{2}$ (in [m s$^{-1}$]) of the mass point robot, namely
\begin{equation}
    a_{t}=\begin{bmatrix}
        a_{x,t}\\
        a_{y,t}
    \end{bmatrix},~~~q_{t}=\begin{bmatrix}
        q_{x,t}\\
        q_{y,t}
    \end{bmatrix}.
\end{equation}
Meanwhile, $F_t$ is given by
\begin{equation}
    F_t=\bar{a}-a_t+u_t,
\end{equation}
where the first term is a pre-stabilizing proportional controller steering the robot to reach the target position $\bar{a}= \boldsymbol{0}$, while $u_t \in \mathbb{R}^{2}$ is the performance-boosting input to be designed with ADMM-PB. This input is designed to make the robot reach the equilibrium point $(\bar{x},\bar{u})=(\boldsymbol{0},\boldsymbol{0})$ faster, while favoring obstacle avoidance and reducing violation of the following robot's velocity constraint
\begin{equation}\label{eq:speed_constraints}
    -\begin{bmatrix}
     0.5\\
     0.5
    \end{bmatrix}\leq q_t\leq  \begin{bmatrix}
     0.5\\
     0.5
    \end{bmatrix},
\end{equation}
while considering the following boosting objective:
\begin{equation}\label{eq:example_loss_robot}
    L(x_{[0,T]}^{s},u_{[0,T]}^{s})=\underbrace{\|x_{[0,T]}^{s}\|_{Q}^{2}\!+\!\|u_{[0,T]}^{s}\|_{R}^{2}}_{=L_{\mathrm{LQ}}(x_{[0,T]}^{s},u_{[0,T]}^{s})}+\alpha L_{\mathrm{ca}}(x_{[0,T]}^{s}).
\end{equation}
We select as weights for the LQ term as $Q=I$, $R=0.1I$ and set $\alpha$ to 10, with the collision avoidance loss (see \cite[Appendix A]{furieri2024learning})
\begin{equation}
    L_{\mathrm{ca}}(x_{[0,T]}^{s})\!=\!\begin{cases}
        \|a_t\!-\!a^{\mathrm{obs}}\|_{2}^{2}\!+\!\nu, ~&\mbox{if } \|a_t\!-\!a^{\mathrm{obs}}\|_{2}^{2} \!\leq\! 1.1 r,\\
        0,~&\mbox{otherwise},
    \end{cases}
\end{equation}
where $r=0.75$ [m] is the sum of the radius of the obstacle and the robot and $\nu=0.001$. Note that this last term penalizes the proximity of the robot to the obstacle when the distance between the two becomes \textquotedblleft unsafe\textquotedblright, with the safety distance here set to $1.1\cdot r$. 

\begin{table}[!tb]
    \centering
     \caption{Hyperparameters of ADMM-PB.}
    \label{tab:ADMM_param}
    \begin{tabular}{cccccccccc}
        $\epsilon^{\mathrm{abs}}$ \hspace*{-.2cm} & \hspace*{-.2cm}$\epsilon^{\mathrm{rel}}$ \hspace*{-.2cm}&\hspace*{-.2cm} $\tau^{\mathrm{inc}}$ \hspace*{-.2cm}&\hspace*{-.2cm} $\tau^{\mathrm{dec}}$ \hspace*{-.2cm}&\hspace*{-.2cm} $\mu$ \hspace*{-.2cm}&\hspace*{-.2cm} $\rho^{(0)}$ \hspace*{-.2cm}&\hspace*{-.2cm} $\eta^{(0)}$ \hspace*{-.2cm}&\hspace*{-.2cm} $\gamma$ \hspace*{-.2cm}&\hspace*{-.2cm} $\mathcal{J}$ \hspace*{-.2cm}&\hspace*{-.2cm} $\bar{\eta}$\\
        \hline
         10$^{-4}$ & 10$^{-4}$ & 2 & 0.5 & 10 & 0.5 & 10$^{-3}$ & 0.5 & 50 & 10$^{-6}$\\
         \hline
    \end{tabular}
\end{table}
\begin{figure}
\centering
    \includegraphics[scale=1]{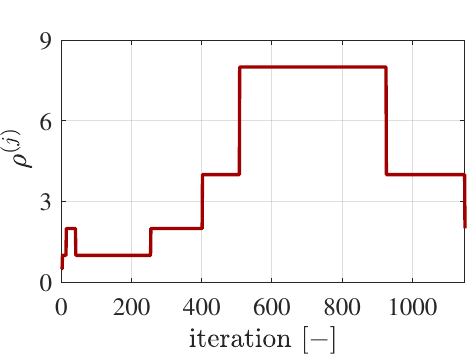} 
    \caption{Evolution of ADMM's step size consequent to the update rules in \eqref{eq:update_rho} and \eqref{eq:step_size_choice}.}
    \label{fig:step_size}
\end{figure}
To parametrize the operator $\mathbfcal{M}$ generating the boosting input as in \eqref{eq:parameterization} we use a Recurrent Equilibrium Network (REN), with hyperbolic tangent activation functions, a 4-dimensional input and 4-dimensional internal state, with the latter initialized at $\boldsymbol{0}$. To train $\mathbfcal{M}$ with ADMM-PB, we consider $S=8$ scenarios with an horizon of $T=249$ steps, initializing the parameters of the REN by drawing $\theta^{(0)}$ at random from a Gaussian distribution with zero mean and variance $0.01I$. The ADMM-PB step in \eqref{eq:ADMM_PB_step1} is carried out considering $6$ epochs for the full batch optimization, using Adam \citep{diederik2014adam} with default parameters, but considering the decaying step size introduced in \eqref{eq:step_size_choice}. The remaining parameters of ADMM-PB using the termination and step updates introduced in Section~\ref{sec:tuning} are reported in \tablename{~\ref{tab:ADMM_param}}, where the parameters chosen for the update of $\rho$ in \eqref{eq:update_rho} correspond to the ones suggested in \cite[Chapter 3.4.1]{boyd2011distributed}. Note that, in line with the discussion in Section~\ref{sec:tuning}, these choices lead to a progressive increase in the ADMM step size up to iteration $508$. In turn, this allows the primal residual to satisfy the termination condition at the $868$-th iteration, as shown in \figurename{~\ref{fig:step_size}}. After that, $\rho$ is reduced again, enabling the dual residual to also satisfy \eqref{eq:termination} and leading to the ADMM-PB automatic termination after $1150$ iterations. 
In terms of computational effort, this setup requires \(1150\) outer ADMM iterations, each with \(6\) full-batch Adam epochs for the \(\theta\)-update, for a total of $6900$ gradient-based training epochs. Meanwhile, as the hard constraints only affect the velocity components and reduce to separable box projections over the copied trajectories, the projection/dual-update stage acts on\footnote{Since \(S=8\), \(T=249\), \(n=4\), and \(m=2\).} \(12000\) copied state-input entries per outer iteration. Hence, given the relative simplicity of the projection step, ADMM-PB has essentially the same dominant training complexity as the baseline rollout-based PB method, plus a relatively cheap feasibility-enforcement step.

\subsubsection{Comparison with a penalty-based PB baseline}
To assess the effect of the proposed ADMM-based constraint-handling mechanism, we compare ADMM-PB with the closest baseline within the PB framework, namely the original training scheme of~\cite{furieri2024learning}. Therein, constraint satisfaction is encouraged through soft penalty terms in the loss. This choice keeps the controller architecture, training setup, and underlying stability framework unchanged, allowing us to isolate the effect of the proposed constraint-handling strategy during training. By contrast, online constrained-control methods such as MPC, reference governors, or safety filters are not directly comparable with our offline training setting. Likewise, approaches based on Lyapunov or barrier certificates typically rely on additional structural information, e.g., a known control Lyapunov or control barrier function. Since this knowledge is not assumed to be available here, any comparison with them would ultimately be unfair.

This comparison is performed by training a REN with the same structure as above via Adam\footnote{In this case, the gradient descent step is fixed to the value of $\eta^{(0)}$ indicated in \tablename{~\ref{tab:ADMM_param}}.} over $E=6900$ epochs, so as to match the total number of gradient-based epochs used within the ADMM routine. To promote constraint satisfaction within the baseline approach, its boosting loss corresponds to \eqref{eq:example_loss_robot} augmented with two additional barrier-inspired penalty terms for each Euclidean coordinate, i.e.,
\begin{subequations}\label{eq:CBF_induced}
    \begin{align}
        & L_{\mathrm{min}}(q_{\xi,[0,T]}^{s})\!=\!\omega\sum_{t=0}^{T-1}\!\left[\max\{0,\!(1\!\!-\!\zeta)\bar{\Delta}_{\xi,t}\!-\!\bar{\Delta}_{\xi,t+1}\}\!\right]\!, \\
        & L_{\mathrm{max}}(q_{\xi,[0,T]}^{s})\!=\!\omega\!\sum_{t=0}^{T-1}\!\left[\max\{0,\!(1\!-\!\zeta)\underline{\Delta}_{\xi,t}\!-\!\underline{\Delta}_{\xi,t+1}\}\!\right]\!,
    \end{align}    
\end{subequations}
where $\bar{\Delta}_{\xi,t\!}\!=\!q_{\xi,t}^{s}\!+0.5$, $\underline{\Delta}_{\xi,t\!}\!=\!0.5-q_{\xi,t}^{s}$, and $\xi$ is a placeholder for $x$ and $y$. In our analysis, we consider $50$ testing scenarios, varying the hyperparameter $\omega$ within the interval $[1,10^{6}]$, fixing $\zeta=0.2$. 

In this comparison, we consider four performance indicators.
First, we look at performance through the average values taken by the LQ term $L_{\mathrm{LQ}}(\cdot)$ and the collision avoidance loss $L_{\mathrm{ca}}(\cdot)$ in \eqref{eq:example_loss_robot} over the $50$ testing scenarios, indicated as $\bar{L}_{\mathrm{LQ}}$ and $\bar{L}_{\mathrm{ca}}$, respectively. To assess the smoothness of training, we also consider the variation of the training loss across epochs 
\begin{equation}
\Delta \tilde{L}=\sum_{i=1}^{E-1}|\tilde{L}_{i}-\tilde{L}_{i-1}|.   
\end{equation}
Here, $\tilde{L}_{i}$ denotes the average training loss at epoch $i$ over the $8$ training scenarios: for ADMM-PB, this is the loss in \eqref{eq:example_loss_robot}, while for penalty-based PB it is the corresponding penalized loss obtained by augmenting \eqref{eq:example_loss_robot} with the soft constraint-penalty terms in \eqref{eq:CBF_induced}.
In addition, to assess performance in terms of constraint violations, we introduce the following indicator
\begin{subequations}\label{eq:indicator_violations}
    \begin{equation}
    V=\sum_{s=1}^{50}\sum_{t=0}^{249}\left[v_{x,t}^{s}+v_{y,t}^{s}\right],
    \end{equation}
where
\begin{equation}
    v_{\xi,t}^{s}=\begin{cases}
        \|\bar{\Delta}_{\xi,t}\|_{2}^{2},~~\mbox{if }~ q_{\xi,t}^{s} < -0.5,\\
        \|\underline{\Delta}_{\xi,t}\|_{2}^{2},~~\mbox{if }~ q_{\xi,t}^{s} > 0.5,\\
        0, \qquad \qquad~\mbox{otherwise,}
    \end{cases}
\end{equation}
and $\xi$ is once again a placeholder for $x$ and $y$. 
\end{subequations}
\begin{table}[!tb]
    \centering
    \caption{ADMM-PB (Algo.~\ref{algo:1}) \emph{vs} penalty-based PB baseline proposed in~\cite{furieri2024learning}: performance indicators.}
    \label{tab:indicators}
\scalebox{0.95}{
    \begin{tabular}{|c|c|c|cccc|}
            \multicolumn{2}{c}{} & \multicolumn{1}{c}{Training} &\multicolumn{4}{c}{Testing} \hspace{-.1cm}\\
            \cline{2-7}
         \multicolumn{1}{c|}{\rule{0pt}{9pt}\rule[-4pt]{0pt}{0pt}} &   \begin{tabular}{c}$\omega$\end{tabular} & $\Delta \tilde{L} \cdot10^5$ & $\bar{L}_{\mathrm{LQ}}$ & $\bar{L}_{\mathrm{ca}}$ & $V$ & $V\!\cdot\! \bar{L}_{\mathrm{LQ}}$ \\
         \hline
         Algorithm \ref{algo:1} & -  & \textbf{\textcolor{green!30!black}{0.7}} & 469.8 & 53.9 & \textbf{\textcolor{green!30!black}{0.39}} & \textbf{\textcolor{green!30!black}{183.2}} \\
         \hline
         \multirow{7}{*}{penalty-based PB baseline} & 10$^{0}$ & 1.1 & \textbf{\textcolor{green!30!black}{241.8}} & 8.6 & \textbf{\textcolor{red!40!black}{1218.3}} & \textbf{\textcolor{red!40!black}{294611.0}} \\
         &  10$^{1}$ & 1.7 & 252.8 & \textbf{\textcolor{green!30!black}{6.0}}  & 902.8 & 228228.6 \\
         &  10$^{2}$ & 4.8 & 376.6 & 77.7 & 18.93 & 7129.0 \\
         &  10$^{3}$ & 1.8 & 413.9 & 21.8 & 5.56 & 2301.3 \\
         &  10$^{4}$ & 3.5 & 456.9 & 15.4 & 5.20 & 2375.9 \\
         &  10$^{5}$ & 40 & 486.6 & 79.8 & 2.99 & 1454.9 \\
         &  10$^{6}$ & 130 & \textbf{\textcolor{red!40!black}{512.0}} & 351.9 & 0.75 & 384.0 \\
         \hline 
    \end{tabular}}
\end{table}
The values of these indicators achieved by ADMM-PB and by the penalty-based baseline for different values of $\omega$ in \eqref{eq:CBF_induced} are reported in \tablename{~\ref{tab:indicators}}. We also include the product $V \cdot \bar{L}_{\mathrm{LQ}}$ to provide a compact indication, of the trade-off between performance and constraint violation.
The results in \tablename{~\ref{tab:indicators}} highlight a practical limitation of penalty-based constraint handling. Indeed, choosing of the weight $\omega$ is all but trivial, with different values potentially leading to markedly different closed-loop behaviors, ranging from aggressive performance-oriented solutions with large violations to more cautious trajectories with degraded performance. Moreover, as shown in \tablename{~\ref{tab:indicators}}, ADMM-PB leads to smoother training than the baseline approach, while different patterns emerge in terms of testing performance. Looking at the average LQ loss $\bar{L}_{\mathrm{LQ}}$, ADMM-PB becomes closer to the baseline as $\omega$ increases. This is consistent with the role of the penalty term in \eqref{eq:CBF_induced}. Indeed, larger values of $\omega$ make the baseline more conservative, so that the robot tends to reach the origin more slowly.
As a result, for smaller values of $\omega$, the baseline reaches the origin faster than ADMM-PB, but typically at the expense of larger constraint violations. This behavior is illustrated in \figurename{~\ref{fig:plane_traj}}, where we compare the trajectories over $50$ testing scenarios using the baseline with\footnote{With this comparison, we consider a case for the baseline that achieves a similar $\bar{L}_{\mathrm{ca}}$ to ADMM-PB.} $\omega=10^{2}$. Note that, after the initial transient, the trajectories of the robot across the testing scenarios tend to be more similar to each other when using ADMM-PB. 
\begin{figure}
\centering
\begin{tabular}{cc}
    \subfigure{\includegraphics[scale=1]{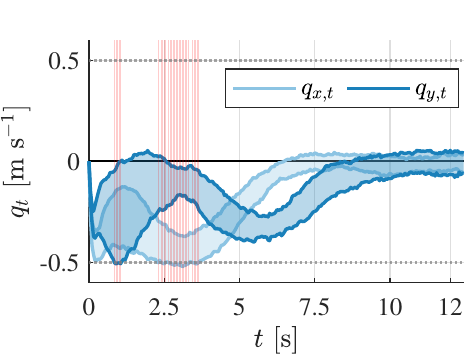}} \hspace*{-.35cm}&\hspace*{-.15cm}  \subfigure{\includegraphics[scale=1]{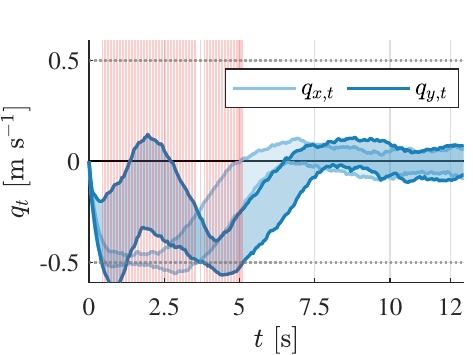}}\\
    (a) ADMM-PB & (b) Baseline for $\omega=10^{2}$
\end{tabular} 
\caption{ADMM-PB \emph{vs} penalty-based PB baseline proposed in~\cite{furieri2024learning}: speed of the robots over time for both Euclidean coordinates $x$ and $y$. The dashed lines indicate the speed constraints in \eqref{eq:speed_constraints}, while the bold colored lines are the minimum and maximum speed achieved by the robot at each time across the $50$ testing scenarios. The red vertical lines are the instants at which constraint violations occur.} \label{fig:speed_traj}
\end{figure}

Nevertheless, the lower $\omega$ gets, the more the penalty-based PB baseline tends to push performance and violate constraints, as clear from the results achieved when looking at $V$ for $\omega=1$ in \tablename{~\ref{tab:indicators}}. This result is further reinforced by the product $V\cdot\bar{L}_{\mathrm{LQ}}$ shown in the same table, indicating that the trade-off between performance and constraint satisfaction is in favor of ADMM-PB. The difference in terms of constraint violation between ADMM-PB and the penalty-based PB baseline for $\omega=10^{2}$ can be further visualized in \figurename{~\ref{fig:speed_traj}}, where we compare the minimum and maximum speed achieved by the robot across the $50$ testing scenarios (the shaded area highlights the speed range of the robot). From this figure, it is clear that the baseline approach pushes performance, requiring an initial velocity that hinges more in the proximity of the imposed lower bound, thereby leading to a higher $V$ than the proposed approach. On the contrary, the speed trajectories resulting from the use of ADMM-PB tend to be more cautious, reducing constraints' violation at the price of more conservative performance.
Note that, although ADMM-PB substantially reduces constraint violations, it does not lead to violation-free closed-loop tests. Since disturbances have here unbounded support, achieving hard constraint satisfaction in the testing phase would in fact require robustification under suitable finite-support or distributional assumptions on the noise, which is beyond the scope of this work.
%
Lastly, by looking again at \tablename{~\ref{tab:indicators}}, it can be noticed that neither ADMM-PB nor the baseline result in $\bar{L}_{\mathrm{ca}}=0$. This result implies that the robot approaches the obstacle beyond the safety distance. At the same time, this does not imply that the robot will collide with the obstacle, as all tests, apart from the one featuring $\omega=10^{6}$, are collision-free. 

\subsubsection{Sensitivity analysis for the hyperparameters}
We now evaluate the sensitivity of ADMM-PB to the choice of its hyperparameters by employing Optuna\footnote{Optuna is a sensitivity analysis framework that operates by sampling the hyperparameter space in a Bayesian fashion to minimize a user-provided metric $L_{\mathrm{hyp}}\in\mathbb{R}$.}~\citep{akiba2019optuna}. We specifically use the primal and dual residuals at the hundredth ADMM iteration, i.e.,
\begin{equation}\label{eq:optuna_metric}
    L_{\mathrm{hyp}}= \|r^{(100)}\|_{2}+\|\delta^{(100)}\|_{2},
\end{equation}
to guide the Bayesian exploration of the parameter space, investigating the importance of the hyperparameters appearing in the update rules in expressions \eqref{eq:update_rho}-\eqref{eq:step_size_choice}. In doing so, we limit the hyperparameter exploration space to
\begin{equation}\label{eq:bound_optuna}
        \tau^{\mathrm{inc}} \in [1.5, 3.0], ~ \tau^{\mathrm{dec}} \in \left[\frac{1}{3}, \frac{2}{3}\right],~\mu \in [5, 15],~ \rho^{(0)} \in [0.1, 10], ~ \eta^{(0)} \in[10^{-5},10^{-2}], ~ \gamma \in [0.5, 0.9], ~\mathcal{J} \in [20, 100],    
\end{equation}
while keeping the lower threshold of the learning rate $\bar{\eta}$ constant. 
In addition, we use log-scale sampling for the initial step-size $\rho^{(0)}$ and learning rate $\eta^{(0)}$.

After $107$ samples, Optuna established the impact of varying each hyperparameter on minimizing \eqref{eq:optuna_metric}, which is reported in Figure \ref{fig:optuna_analysis}(a). From this, we observe that modifying the parameter $\mu$, which dictates the scaling of the primal-dual difference depending on how $\rho$ is updated through \eqref{eq:update_rho}, has essentially no effect on minimizing the residuals. At the same time, the main tuning knobs to achieve small residuals in our benchmark appear to be $\tau^{\mathrm{inc}}$ for the ADMM step-size update in \eqref{eq:update_rho} and $\gamma$ for the learning-rate update.

We display in Figure \ref{fig:optuna_analysis}(b) the violin plot for the hyperparameters tested by Optuna together with the corresponding achieved value of $L_{\mathrm{hyp}}$, where the hyperparameters have been normalized according to their minimum and maximum extracted values. 
From this, we can understand that values close to the lower bound in \eqref{eq:bound_optuna} for the initial step-size, learning rate, and $\tau^{\mathrm{dec}}$ favor a lower value of the tested metric, and that the same applies to intermediate values of the scheduler decay and high values of its number of iterations. Finally, by combining the interpretations of the results in both figures, we can conclude that the primary parameter to tune for the robot benchmark is $\tau^{\mathrm{inc}}$, due to its broad distribution in the violin plot, and that the other parameter values should be set after fixing it.

\begin{figure}
    \centering
    \begin{minipage}[t]{0.48\textwidth}
        \centering
        \begin{tikzpicture}
        \begin{axis}[
            width=\linewidth,
            height=0.75\linewidth,
            xbar,
            xmin=0,
            xlabel={Percentage \%},
            y dir=reverse,
            symbolic y coords={tauinc, taudec, mu, rho, eta, gamma, J},
            ytick=data,
            yticklabels={
                {$\tau^{\mathrm{inc}}$},
                {$\tau^{\mathrm{dec}}$},
                {$\mu$},
                {$\rho^{(0)}$},
                {$\eta^{(0)}$},
                {$\gamma$},
                {$\mathcal{J}$}
            },
            nodes near coords,
            nodes near coords align={horizontal},
            axis y line*=left,
            axis x line*=bottom,
            tick label style={font=\scriptsize},
            label style={font=\scriptsize},
            yticklabel style={font=\scriptsize},
            every node near coord/.append style={font=\scriptsize}
        ]
            \addplot coordinates {
                (33,tauinc)
                (7,taudec)
                (0.01,mu)
                (6,rho)
                (12,eta)
                (29,gamma)
                (13,J)
            };
        \end{axis}
        \end{tikzpicture}

        \vspace{1mm}
        \footnotesize{(a) Relative importance of the tested hyperparameters for the metric in \eqref{eq:optuna_metric}.}
    \end{minipage}
    \hfill
    \begin{minipage}[t]{0.48\textwidth}
        \centering
        \includegraphics[width=\linewidth]{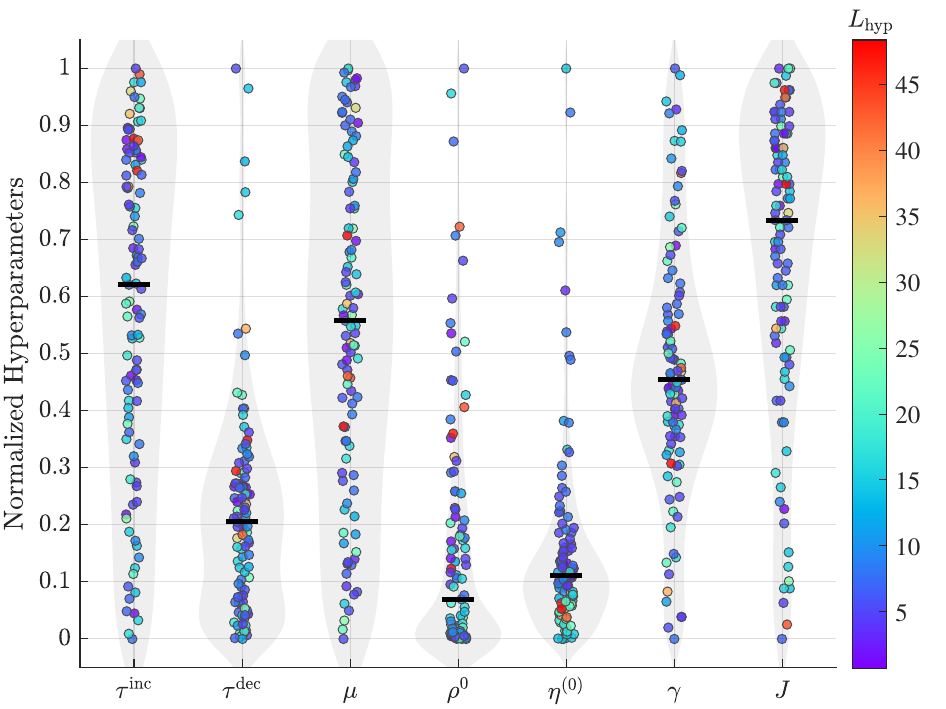}

        \vspace{1mm}
        \footnotesize{(b) Normalized distribution of the 107 hyperparameter samples explored by Optuna.}
    \end{minipage}
    \caption{Optuna-based hyperparameter analysis for ADMM-PB. Panel (a) reports the relative impact of varying each hyperparameter within the ranges in \eqref{eq:bound_optuna} on the metric in \eqref{eq:optuna_metric}. Among the tested hyperparameters, $\tau^{\mathrm{inc}}$ has the largest influence on the metric, while $\mu$ has a negligible effect. Panel (b) shows the normalized distribution of the 107 sampled hyperparameter configurations, where each parameter is scaled using the minimum and maximum extracted values. }
    \label{fig:optuna_analysis}
\end{figure}

\subsection{Constrained control of a three-tank system}
\begin{figure*}
    \centering
\tikzset{every picture/.style={line width=0.75pt}} 
\begin{tikzpicture}[x=0.75pt,y=0.75pt,yscale=-1,xscale=1]
\draw [line width=1.5]    (58,101) -- (78,101) ;
\draw [line width=1.5]    (116,101) -- (125,101) ;
\draw [line width=1.5]    (87,101) -- (107,101.5) ;
\draw [line width=1.5]    (87,100) -- (87,107) ;
\draw [line width=1.5]    (78,100) -- (78,107) ;
\draw [color={rgb, 255:red, 74; green, 144; blue, 226 }  ,draw opacity=1 ][line width=0.75]    (58,76) .. controls (111,70) and (89,78) .. (125,76) ;
\draw [line width=1.5]    (116,100) -- (116,107) ;
\draw [line width=1.5]    (106,101) -- (106,114.5) ;
\draw [line width=1.5]    (147,114.5) -- (147,156) ;
\draw [line width=1.5]    (146,157) -- (159,157) ;
\draw [line width=1.5]    (200,157) -- (209,157) ;
\draw [line width=1.5]    (171,157) -- (191,157.5) ;
\draw [line width=1.5]    (168,156) -- (168,163) ;
\draw [line width=1.5]    (159,156) -- (159,163) ;
\draw [color={rgb, 255:red, 74; green, 144; blue, 226 }  ,draw opacity=1 ][line width=0.75]    (148,131) .. controls (201,125) and (172,133) .. (208,131) ;
\draw [line width=1.5]    (200,157) -- (200,164) ;
\draw    (120,23) -- (161,23) ;
\draw   (161,23) .. controls (161,18.58) and (164.58,15) .. (169,15) .. controls (173.42,15) and (177,18.58) .. (177,23) .. controls (177,27.42) and (173.42,31) .. (169,31) .. controls (164.58,31) and (161,27.42) .. (161,23) -- cycle ;
\draw   (161,23) -- (173.67,16.67) -- (173.51,29.64) -- cycle ;
\draw    (120,23) -- (120,86.5) ;
\draw    (177,23) -- (286,23.5) ;
\draw [line width=1.5]    (147,115.5) -- (105,115.5) ;
\draw [line width=1.5]    (157,107) -- (115,107) ;
\draw [line width=1.5]    (208,104) -- (208,156) ;
\draw [line width=1.5]    (200,157) -- (209,157) ;
\draw [line width=1.5]    (169,157) -- (189,157.5) ;
\draw [line width=1.5]    (200,156) -- (200,163) ;
\draw [line width=1.5]    (190,157) -- (190,170.5) ;
\draw [line width=1.5]    (231,169.5) -- (231,211) ;
\draw [line width=1.5]    (293,159) -- (293,211) ;
\draw [line width=1.5]    (232,210) -- (245,210) ;
\draw [line width=1.5]    (254,210) -- (292,210) ;
\draw [line width=1.5]    (254,209) -- (254,216) ;
\draw [line width=1.5]    (245,209) -- (245,216) ;
\draw [color={rgb, 255:red, 74; green, 144; blue, 226 }  ,draw opacity=1 ][line width=0.75]    (232,186) .. controls (285,180) and (256,188) .. (292,186) ;
\draw [line width=1.5]    (231,170.5) -- (189,170.5) ;
\draw [line width=1.5]    (241,164) -- (199,164) ;
\draw    (286,23.5) -- (286,193.5) ;
\draw    (38.64,23) -- (38.83,68) ;
\draw   (38.83,68) .. controls (43.25,67.98) and (46.85,71.55) .. (46.87,75.97) .. controls (46.88,80.39) and (43.32,83.98) .. (38.9,84) .. controls (34.48,84.02) and (30.88,80.45) .. (30.87,76.03) .. controls (30.85,71.62) and (34.41,68.02) .. (38.83,68) -- cycle ;
\draw   (38.83,68) -- (45.22,80.64) -- (32.24,80.54) -- cycle ;
\draw    (38.9,84) -- (38,243.5) ;
\draw    (62,23) -- (38.64,23) ;
\draw    (62,23) -- (62,86.5) ;
\draw [color={rgb, 255:red, 74; green, 144; blue, 226 }  ,draw opacity=1 ][line width=0.75]    (20,235.75) .. controls (87,225) and (215,243) .. (310,234) ;
\draw [line width=1.5]    (20,250) -- (310,250) ;
\draw [line width=1.5]    (310,219) -- (310,251) ;
\draw [line width=1.5]    (20,220.5) -- (20,251) ;
\draw [line width=1.5]    (58,50) -- (58,102) ;
\draw [line width=1.5]    (125,50) -- (125,102) ;
\draw    (131,74) -- (150,74) ;
\draw    (131,100) -- (150,100) ;
\draw    (140,74) -- (140,100) ;
\draw    (213,130) -- (232,130) ;
\draw    (213,156) -- (232,156) ;
\draw    (222,130) -- (222,156) ;
\draw    (299,184) -- (318,184) ;
\draw    (299,210) -- (318,210) ;
\draw    (308,184) -- (308,210) ;
\draw (70,53) node [anchor=north west][inner sep=0.75pt]   [align=left] {tank 1};
\draw (156,110) node [anchor=north west][inner sep=0.75pt]   [align=left] {tank 2};
\draw (239,166) node [anchor=north west][inner sep=0.75pt]   [align=left] {tank 3};
\draw (145,78.4) node [anchor=north west][inner sep=0.75pt]    {$h_{1}$};
\draw (226,134.4) node [anchor=north west][inner sep=0.75pt]    {$h_{2}$};
\draw (312,188.4) node [anchor=north west][inner sep=0.75pt]    {$h_{3}$};
\draw (13,67.4) node [anchor=north west][inner sep=0.75pt]    {$v$};
\end{tikzpicture}
    \caption{Triple-tank system with recirculation pump.}
    \label{fig:triple-tank}
\end{figure*}

Our second benchmark for ADMM-PB is the three-tank system with recirculating structure analyzed in \cite{massai2025l2ru,massai2024unconstrained}, and shown in Figure~\ref{fig:triple-tank}. Note that, unlike the robot example, the goal of this benchmark is not to emphasize sensitivity to penalty tuning. Instead, we employ it to show that ADMM-PB can be used as a structure-preserving constrained refinement of PB, and to demonstrate how classical PB can provide a reasonable warm start for ADMM-PB.

The dynamics of the three-tank system are described by
\begin{equation}\label{eq:three_tank_dyn}
    x_t =  
x_{t-1} 
 + T_s \left(
\begin{bmatrix} 
-\frac{a_1}{A_1} & 0 & k_1 \frac{a_3}{A_1} \\ 
k_2 \frac{a_1}{A_2} & -\frac{a_2}{A_2} & 0 \\ 
0 & k_3 \frac{a_2}{A_3} & -\frac{a_3}{A_3} 
\end{bmatrix}
\begin{bmatrix} 
\sqrt{2gh_{1, t-1}} \\ 
\sqrt{2gh_{2, t-1}} \\ 
\sqrt{2gh_{3, t-1}} 
\end{bmatrix} + 
\begin{bmatrix} 
\frac{k_c}{A_1} \\ 
0 \\ 
0 
\end{bmatrix} v_{t-1}
\right) + w_{t},
\end{equation}
where $x_t=\begin{bmatrix} h_{1,t} & h_{2,t} & h_{3,t} \end{bmatrix}^\top\in\mathbb{R}^3$~[cm] collects the fluid levels, $v_t$~[V] is the voltage applied to the pump, while the model parameters are the same as in \cite{massai2025l2ru,massai2024unconstrained}. Moreover, the initial condition $x_0$ is sampled from a uniform distribution centered at the equilibrium point $\bar{x}=\begin{bmatrix}551.06&80.97&5.47\end{bmatrix}^\top$, with support widths equal to $20$, $10$, and $1$ for the three tanks, respectively. Meanwhile, for $t\geq 1$, the disturbance $w_t$ is sampled from a uniform distribution centered at $0$ with support widths equal to $1$, $0.1$, and $0.1$. The input voltage is here defined as
\begin{equation*}
    v_t = \bar{v}+ u_t,
\end{equation*}
where $\bar{v}=1$~[V] yields the equilibrium $\bar{x}$ and $u_t$ is the performance boosting component. In this example, we promote feasibility with respect to the constraints
\begin{subequations}
    \begin{align}
    -0.5 \leq u_t &\leq 0.5, \label{eq:constr_input_1_tank}\\
    -0.1\leq\Delta u_t&\leq 0.1,\label{eq:constr_input_2_tank}
\end{align}
\end{subequations}
where $\Delta u_t=u_{t-1} -u_t$ and 
\begin{equation*}
    \Delta u_{[1,T]} = \begin{bmatrix}
        \Delta u_1 ^\top & \Delta u_2 ^\top & \cdots &  \Delta u_T ^\top
    \end{bmatrix}^\top.
\end{equation*}
Meanwhile, as the boosting objective, we use
\begin{equation}\label{eq:example_loss_tank}
    L(x_{[0,T]}^{s},u_{[0,T]}^{s})\!=\!\underbrace{\|x_{[0,T]}^{s}-1_{T+1}\bar{x} \|_{Q}^{2}\!+\!\|u_{[0,T]}^{s}\|_{R}^{2}}_{=L_{\mathrm{LQ}}(x_{[0,T]}^{s},u_{[0,T]}^{s})}+ \underbrace{\|\Delta u_{[1,T]}^{s}\|_{\mathcal{R}}^{2}}_{=L_\Delta(u_{[0,T]}^{s}) },
\end{equation}
with weights $Q=\mathrm{diag}(0.01,0.1,1)$, $R=I$, and $\mathcal{R}=1$, and in addition the following penalty 
\begin{equation}
\label{eq:thank_penalty}
        L_{B}(u_{[0,T]}^{s})\!=\sum_{t=1}^{T-1}\!\left[\max\{0.1, |\Delta u_t | \}\!\right] + \sum_{t=0}^{T-1}\left[\max\{0, u_t -0.5 \} + \max\{0, -0.5 -u_t\}\right]. 
\end{equation} 
Note that the operator $\mathbfcal{M}$ has the same structure as in the robot benchmark.

To achieve the goal of this example, i.e., show how ADMM-PB can refine classical PB solutions and how the latter can be used as reasonable warm starts for ADMM-PB, we perform the following steps. First, we train the PB controller with a penalty-based scheme until convergence. We do so considering penalty weights in the set $\{5\cdot 10^{1},\,5\cdot 10^{2},\,5\cdot 10^{3},\,5\cdot 10^{4},\,5\cdot 10^{5} \}$, while using \eqref{eq:example_loss_tank} augmented with soft penalties on violations of \eqref{eq:constr_input_1_tank}--\eqref{eq:constr_input_2_tank}. We then select the controller achieving the best validation performance, that is with a soft penalty weight of $500$, and use the corresponding REN parameters to warm-start ADMM-PB. The latter is subsequently trained over $S=60$ scenarios with horizon $T=200$, using full-batch updates of $6$ epochs, $\epsilon^{\mathrm{abs}}=10^{-4}$, $\epsilon^{\mathrm{rel}}=10^{-3}$, constant ADMM step-size $\rho=3.75$, and fixed learning rate $\eta=10^{-3}$.

Starting from this initialization, ADMM-PB improves constraint satisfaction without modifying the PB structure nor compromising its stability-by-design properties. In particular, across $20$ testing scenarios, ADMM-PB achieves an approximately $20\%$ reduction in the constraint-related penalty term~\eqref{eq:thank_penalty} with respect to the best penalty-based PB controller selected on the validation set. As is clear from Figure~\ref{fig:three_tank_traj_inputs}, we observe no violations of the input bound \eqref{eq:constr_input_1_tank}. Meanwhile, the tighter constraint on $\Delta u_t$ in \eqref{eq:constr_input_2_tank} is still violated at the beginning of the testing horizon. Yet violations are reduced relative to those attained with the controller used to warm-start ADMM-PB. This improvement is obtained without degrading the control performance, which is slightly improved by about $0.27\%$. These results suggest that, on this benchmark, ADMM-PB can be effectively used as a post-training constrained refinement layer on top of PB, benefiting from and enhancing the baseline procedure.

\begin{figure}
\centering
\begin{tabular}{cc}
    \subfigure{\includegraphics[scale=1]{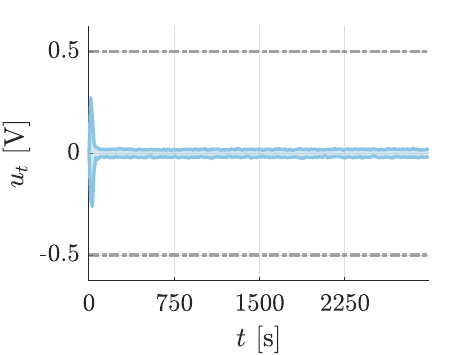}} \hspace*{-.35cm}&\hspace*{-.35cm}  \subfigure{\includegraphics[scale=1]{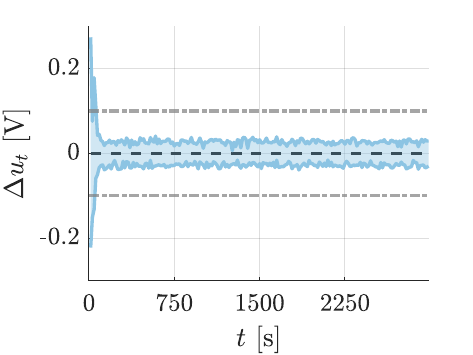}} 
\end{tabular} \caption{Input evolution and its variation achieved with ADMM-PB for $20$ testing scenarios. The dashed gray lines are the imposed constraints.}\label{fig:three_tank_traj_inputs}
\end{figure}

\section{Conclusions}\label{sec:conclusion}
Building on the performance boosting approach proposed in \cite{furieri2024learning}, this work considers the offline training of stable-by-design neural controllers in the presence of prescribed state and input constraints. To tackle this problem, we propose ADMM-PB, an ADMM-inspired training routine that accounts for these constraints via auxiliary trajectory variables and projection steps, without structurally altering the controller used in~\cite{furieri2024learning}. The resulting controller, therefore, remains in the same stable parametrized class, while the training procedure provides a structured mechanism to promote constraint satisfaction. 
Numerical validation on two complementary benchmarks highlights the possible advantages of the proposed strategy, which results in smoother training and a better trade-off between constraint violation reduction and performance boosting than penalty-based constrained training. Our test also shows that ADMM-PB can be used to reduce constraint violations 
starting from a PB controller trained with soft penalties, without modifying the underlying architecture. This result opens up the possibility of using ADMM-PB as a post-training refinement strategy to enhance a baseline performance booster.    


\bibliographystyle{cas-model2-names}

\bibliography{Main.bib}



\end{document}